%% file: main.tex
\documentclass[twocolumn]{aastex63}
\usepackage[utf8]{inputenc}
\usepackage{amsmath}
\usepackage{amsfonts}
\usepackage{amssymb}
\usepackage{graphicx}
\usepackage{grffile}
\usepackage{hyperref}
\usepackage{array}
\usepackage{listings}
\usepackage{comment}
\usepackage{bm}
\usepackage{slashed}
\usepackage{mathtools}
\let\tablenum\relax
\usepackage{siunitx}
\usepackage{multirow}
\usepackage{booktabs}
\usepackage[para]{threeparttable}

\usepackage{float}

\usepackage{ulem}
\usepackage{xcolor}
\input{defs.tex}

\newcommand{\cc}{{\cal C}}
\newcommand{\co}{{\rm col}}

\newcommand{\hto}{{\rm H_2}}

\newcommand{\hd}{{\rm HD}}
\newcommand{\rH}{{\rm H}}

\newcommand{\re}{{\rm e}}
\numberwithin{equation}{section}

\begin{document}
\title{Zero Metallicity with Zero CPU Hours: Masses of the First Stars on the Laptop}
\author{James Gurian}
\affiliation{Institute for Gravitation and the Cosmos, The Pennsylvania State University, University Park, PA 16802, USA}
\affiliation{Department of Astronomy and Astrophysics, The Pennsylvania State University, University Park, PA, 16802, USA}
\affiliation{Perimeter Institute for Theoretical Physics, Waterloo, Ontario, N2L 2Y5, Canada}

\author{Donghui Jeong}
\affiliation{Institute for Gravitation and the Cosmos, The Pennsylvania State University, University Park, PA 16802, USA}
\affiliation{Department of Astronomy and Astrophysics, The Pennsylvania State University, University Park, PA, 16802, USA,
}
\affiliation{School of Physics, Korea Institute for Advanced Study (KIAS), 85 Hoegiro, Dongdaemun-gu, Seoul, 02455, Republic of Korea}

\author{Boyuan Liu}
\affiliation{Institute of Astronomy, University of Cambridge, Madingley Road, Cambridge, CB3 0HA, UK}

\begin{abstract}
We develop an analytic model for the mass of the first stars forming in the center of primordial gas clouds as a function of host halo mass, redshift, and degree of rotation. The model is based on the estimation of key timescales determining the following three processes: the collapse of the gas cloud, the accretion onto the protostellar core, and the radiative feedback of the protostellar core. The final stellar mass is determined by the total mass accreted until the radiative feedback halts the accretion. The analytic estimation, motivated by the result of the full numerical simulations, leads to algebraic expressions allowing an extremely fast execution. Despite its simplicity, the model reproduces the stellar mass scale and its parameter dependences observed in state-of-the-art cosmological zoom-in simulations. This work clarifies the basic physical principles undergirding such numerical treatments and provides a path to efficiently calibrating numerical predictions against eventual observations of the first stars.
\end{abstract}

\section{Introduction}

The primordial universe contained only trace metals from big bang nucleosynthesis \citep[BBN;][]{BBNreview:2007}. The first (Population III, hereafter Pop. III) stars must have formed from this pristine gas, which was cooled principally by atomic and molecular hydrogen. Still, no one has observed a Pop.~III star, so our knowledge of this process comes entirely from simulations and analytic estimates \citep[see][for reviews]{Bromm2004,Bromm:2013, Haemmerl__2020,klessen2023stars}.

Due to the lack of observational comparison and complex physics involved, the Pop.~III star formation process is still an uncertain and active area of research. In the hierarchical structure formation scheme, these stars first formed in the universe when halos massive enough to cool and collapse their gas began to virialize. This process began at redshift $z\sim 30$ and in halos of masses $\sim 10^5$--$10^6 \, \rm M_\odot$ \citep{Tegmark1997,Bromm2004}. The Pop. III stars ended the dark ages and began the reionization and metal enrichment of the universe, and the turn-on of Pop. III stars marks the epoch when, for the first time since BBN, nuclear physics becomes relevant in the universe.

In the standard $\Lambda$CDM cosmology, the characteristic abundance of the mini-halos which could host Pop.~III stars is $\sim 100 \,\rm cMpc^{-3}$ at redshift $20-30$ \citep{Yoshida/etal:2006}. Here, cMpc is the comoving megaparsec. Hence, to develop a statistical sample of tens to hundreds of primordial star-forming clouds, simulations must consider a volume of $\sim \rm cMpc^3$, while the characteristic scale of the proto-stellar accretion disk is $\sim 100\rm AU$. This corresponds to a dynamic range of some ten orders of magnitude, which requires zoom-in simulations. The zoom-in simulations beginning from cosmological initial conditions \citep{Greif2011,Hirano2015, Stacy2016,Susa2014} reveal that Pop.~III stars form in small star clusters, with relatively more massive stars (between a few tens and a few thousand solar mass) forming in the center of the collapsing cloud and several lower mass companions which originate from fragmentation in the star-forming disk, as illustrated in Figure 1 of \cite{Liu2020}. 

Studying the primordial star-cluster systems' formation and evolution demands three-dimensional radiation hydrodynamics simulations in a dense, optically thick environment. Due to computational limits, simulations often do not evolve these primordial clusters to their full maturity, when accretion is shut off by (proto-)stellar feedback, and can only report on the trends observed during the simulation time span. Moreover, since these simulations typically select the first star-forming cloud to form in the cosmological volume as the zoom-in region, they are subject to sampling bias and variance \citep{stacy_constraining_2013}.

On the other hand, tracking the evolution of a single proto-star in a collapsing cloud including accretion, feedback, and radiative transfer is numerically tractable \citep{Hosokawa2011, hosokawa_rapidly_2012, hirano_one_2014, Hosokawa2016, Sharda_2020, Sharda_2021, Latif_2022}. \cite{hirano_one_2014} attempted to characterize the population of primordial stars by beginning with a cosmological volume of $(2 \, {\rm Mpc}/h)^3$, resolving $\sim 100$ individual star-forming clouds. After its formation, the density of each cloud was azimuthally averaged, then used to initialize 2D radiation hydrodynamics simulations, sidestepping the most computationally demanding part of the problem. Note that the azimuthal averaging {\it smears out} any possible small-scale fragmentation and does not resolve the formation and evolution of stellar multiples. However, it allows the central, most massive star in each halo to be evolved until accretion is shut off by feedback. \cite{hirano_one_2014} found that in their simulations the mass of the central star in each gravitationally unstable cloud is tightly correlated with the collapse timescale of the star-forming cloud. They also found that a long collapse timescale (associated with a low halo mass) allows for the production of $\hd$ molecules, permitting cooling to lower temperatures and promoting the formation of low-mass stars. The similar study of \cite{Susa2014} did not include $\hd$ chemistry, and found no such correlations. 

Here, we expand upon \citet{hirano_one_2014} by deriving the relationships revealed by the simulations. We show that a simple analytic model based on algebraic timescale arguments can capture the formation of the most massive Pop.~III star in the center of a collapsing gas cloud. In particular, this model reproduces the stellar-mass distribution of the sophisticated numerical treatment of \cite{hirano_one_2014}. In the process, we clarify the most important physics underlying Pop. III star formation. 

Such an analytic method provides a theoretical ``handle'' on the Pop. III stellar-mass as a function of the mass and formation redshift of the hosting halo and the rotation parameter. As observations of stellar populations at high redshift become available using, for example, the JWST satellite, such a handle will be necessary to efficiently connect the simulated universe to the physical Universe and to extract cosmological information from the Pop.~III observables. We draw upon previous analytic studies \citep{mckee_formation_2008, Stahler1986}, generalizing those results to include the effect of the host halo mass and formation redshift on the final stellar mass while wherever possible simplifying the arguments to simple, algebraic relations. 

Finally, we contrast the Pop.~III star modeling presented here to the modeling of the later generations (Population I and II). The project of deriving the mass of Pop.~III stars {\it ab initio} is saved from hopelessness only by the simplicity of their environment. These stars form from the first clouds to become gravitationally unstable as halos which exceed the cosmological Jeans mass begin to form. Because the first stars arise from the first baryonic structures to depart from the underlying dark matter distribution, there is reason to believe that their properties can be inferred from the well-understood evolution of the dark matter distribution and basic timescale arguments. For this reason, environmental effects can be described in terms of a small number of parameters at the scale of the halo or star-forming cloud. That is, deriving the population statistics of the first stars does not require resolving the detailed gas physics, feedback, and radiative transfer in the star-forming clouds over a cosmological volume because the initial conditions can be accurately described using only a few parameters. We defer a detailed study of the enviromental effects such as the Lyman-Werner radiation \citep[e.g.][]{Nebrin_2023} and the relative velocity between the baryons and dark matter \citep[e.g.][]{Nakazato_2022} to a future study. 

 The paper is organized as follows. In \refsec{Timescales} we present a basic description of the dynamics of the collapsing gas, which informs our determination of the mass of the gravitationally unstable cloud. In \refsec{GasChem} we describe the chemical-thermal network for the primordial gas and compare our results with standard one-zone calculations. In \refsec{MStar} we develop relations governing the growth of the central star and evaluate the final stellar mass over a range of environmental conditions. We conclude with a discussion of the context of this work and possible future avenues for research. 

\section{Collapse Dynamics and Cloud Mass}
\label{sec:Timescales}

The evolution of a primordial gas cloud can be broadly understood through the relations between the following timescales:
\begin{itemize}
    \item The free-fall timescale $t_{ff}= \sqrt{\frac{3\pi}{32G\rho}}$, where $\rho$ is the total, dark matter and baryon, density, which is the time for a test particle accelerated by gravity to fall to the center of the cloud.
    \item The sound crossing timescale $t_s = r/c_s$ where $r$ is the radius of the cloud and $c_s = \sqrt{\frac{\gamma k_BT}{\mu m_P}}$ is the adiabatic sound speed with the adiabatic index $\gamma$, temperature $T$ and mean molecular weight $\mu$ of the cloud, which is the time for a pressure wave to propagate across the cloud.
    \item The cooling timescale $t_{\cc} = {\cal E}_{th}/\cc$, where ${\cal E}_{th}$ is the thermal energy density of the cloud ($\rm [energy][volume]^{-1}$) and $\cc$, the volumetric cooling rate ($\rm [energy][time]^{-1}[volume]^{-1}$), which is  the time for a gas cloud to lose its thermal energy by cooling. Note that the cooling rate depends on the density, composition, and temperature of the cloud. 
    \item The $\hto$ formation timescale $t_{\hto}$, which is the time for enough molecular hydrogen to be produced to cool the cloud. Although we use $t_{\hto}$ only in qualitative discussions in this section, we will provide a more quantitative definition of $t_\hto$ in \refsec{mcloud}.
    \item The collapse timescale $t_{\rm col}$ is an e-folding time for the cloud's density, which will be determined from the preceding timescales. 
\end{itemize}

Here's the qualitative sketch of how these timescales interplay to determine the collapsing gas cloud's mass.

The gas in a primordial halo is initially near virial equilibrium, which is equivalent (up to factors of order unity) to the boundary ($t_{ff}=t_s$) of the Jeans stability condition $t_{ff} > t_s$. This stability condition can also be translated to the Jeans mass,
\begin{align}
   M_J &= \frac{4 \pi\rho c_s^3}{3t_{ff}^3} \\
   &\approx 1.44\left(\frac{ k_B T}{\mu m_P G }\right)^{3/2}\rho^{-1/2}, \label{eq:mj}
\end{align}
with $k_B$ Boltzmann's constant, $T$ the gas temperature, $\mu$ the mean molecular weight, $m_P$ the proton mass, $G$ Newton's constant, and $\rho$ the gas density. In the second line, we have taken $\gamma = 5/3$ for a monatomic ideal gas. Note that different derivations of the Jeans criterion yield different values for the order-unity prefactor, here $1.44$. 
Setting $M_J$ approximately equal to $M_H$, the halo mass, determines the virial temperature $T$ of the halo at its formation $\rho\simeq 178\bar{\rho}_{\rm m}(z)$ , where $\bar{\rho}_{\rm m(z)}$ is the background matter (dark matter and baryon) density at redshift $z$. 

Collapse from this marginally stable configuration at the virial equilibrium requires cooling, which increases the sound crossing time by reducing the temperature. The dominant coolant of primordial clouds, which consist of pristine gas and whose  virial temperature is less than $\sim10^4$ K, is molecular hydrogen. The cosmological molecular hydrogen fraction $x_{\hto,0} \equiv n_{\hto}/n_{\rm H}$ that sets the cloud's initial molecular hydrogen abundance is insufficient to cool the cloud. Therefore, cooling occurs only after enough molecular hydrogen piles up, and the collapse timescale is ultimately determined by the $\hto$ production timescale $t_\hto$.

The ability of pressure to inhibit collapse depends not only on the temperature but also on the mass of the gas cloud: pressure support can be overcome either by cooling (reducing $M_J$) or by growing more massive ($M_{\rm cloud}$ exceeding fixed $M_J$). This fact is responsible for the ``loitering phase'' \citep{Bromm2002}, which is a pause in the condensation of the gas at the density where cooling becomes inefficient. At a critical density  $n_{\rm crit}\sim 10^{3} \, \rm cm^{-3}$, collisional de-excitation reduces the efficiency of $\hto$ cooling. Beyond this density, collapse can continue only once enough gas particles have condensed into a cloud at this ``loitering'' density to exceed the corresponding Jeans mass. Molecular hydrogen alone can cool the gas to $\sim 200 \, \rm K$ at $n_{\rm crit}$ (see, for example, \reffig{h2onez}). The Jeans mass at this temperature and $n_{\rm crit}$ is $\sim 1000\, M_\odot$, which simulations confirm is the approximate mass scale of the gravitationally unstable clouds, when the cloud is cooled by $\hto$ alone \citep{Bromm2002}.

However, if appreciable $\hd$ is formed, the gas can reach temperatures as low as $\sim 50 \, \rm K$, which suppresses the Jeans mass at the critical density and hence the mass of the star-forming cloud. At temperatures above $\sim 500 \, \rm K$, $\hd$ is efficiently converted into $\hto$, so chemical fractionation of $\hd$ begins at $T=500\ \rm K$. When collapse occurs rapidly (on the free-fall timescale $t_{ff}$, for example) the time between $T = 500\, \rm K$ and the end of the loitering phase at $T = 200 \, \rm K$ is too brief to form significant $\hd$. However, if the collapse occurs more slowly, with a long enough $\hto$ production timescale $t_{\hto}$, then appreciable $\hd$ can form, allowing for further cooling and hence a lower cloud mass. The mass of the collapsing cloud in turn determines the mass scale of the central protostar. Environmental effects including mergers and shocks can influence the efficiency of $\hd$ cooling \citep{magnus2023formation, Johnson2006}. In this work we consider only the case where $\hd$ is formed due to a delay in the collapse according to the intrinsic properties of the halo: its mass and its redshift, which are assumed to capture all environmental effects.

\section{Gas Chemistry}
\label{sec:GasChem}

Our analytical estimation is based on a so-called ``one-zone'' calculation, which follows the chemical-thermal evolution of a uniform density parcel of gas evolving in a free-fall timescale. For the typical density profile of primordial gas clouds, where the density profile $\rho(r)$ decreases as a function of radius, the free-fall timescale of the inner part of the cloud is much shorter than the outskirts. The inner part therefore undergoes more time-scales after a fixed elapsed time, and one can think of this inner part as the later evolutionary stage of the one-zone calculation. We therefore expect that the time evolution of one-zone calculation describes the radial profile of the cloud at a fixed time, and indeed the radial profile of full three-dimensional simulations shows a clear correspondence with the results of one-zone calculations \citep{Yoshida/etal:2006}. 

The thermal evolution of such a gas parcel subject to radiative cooling and adiabatic heating is described by:
\begin{equation}
    \frac{dT}{dt} = (\gamma -1)\left(\frac{ \dot n}{n}T-\frac{{\cal C}(T,{\vec n})}{k_B n} \right),
    \label{eq:Tevol}
\end{equation}
where $T$ is the temperature, $\gamma$ the adiabatic index, $k_B$ Boltzmann's constant, and $\vec{n}$ the number densities of the various species and $n$ the total nucleon density (i.e.~including helium). 
The first term in the parentheses describes compressional heating due to adiabatic collapse $TV^{\gamma-1}=$const. 

One-zone calculations do not self-consistently solve for gravity: the density evolution $\dot n$ must be independently specified, and we use a generic parameterization in terms of the collapse timescale $t_{\rm col}$ as 
\begin{align}
    \frac{d n}{dt} &= -3 n \frac{\dot r}{r} 
    \approx \frac{n}{t_{\co}(n)},
    \label{eq:nevol} 
\end{align}
which comports with the definition of $t_{\rm col}$. It is possible to numerically integrate Eq.~\ref{eq:Tevol} along with the equations for the chemical abundances. Here, we  will simplify the problem further by reducing Eq.~\ref{eq:Tevol} to an algebraic equation with a clear physical interpretation. Because one-zone calculations are already inexpensive (less than a few seconds on modern consumer hardware), the computational speedup is unlikely to be important. However, this derivation emphasizes the simple physics which determine the density-temperature relationship of the collapsing gas. We begin by rewriting the temperature evolution in terms of the relevant timescales: the collapse timescale $t_{\rm col}$ and the cooling timescale $t_{\cal C}$.
\begin{equation}
    \frac{d \log T}{d \log n}= (\gamma -1)\left[
    1 -\frac{t_{\co}(n)}{(\gamma-1)t_{\cc}({\vec n},T)}\right].
    \label{eq:evol}
\end{equation}
As expected, for $t_{\co} \ll t_{\cc}$  we regain the adiabatic heating $T\propto n^{(\gamma-1)}$, while for $t_{\co} \gg t_\cc$ the gas cools. 

The temperature and density evolutions described in \refeqs{Tevol}{nevol} are self regulatory: if $t_\co$ differs greatly from $t_\cc$ at any point, the system will evolve towards $t_{\co} \sim t_\cc$ and vice versa. Specifically, the solution 
\begin{equation}
    t_{\co}(n) = (\gamma-1) t_{\cc}({\vec n},T)
    \label{eq:att}
\end{equation} 
is an attractor. In more physical language, the collapsing gas will evolve towards the equilibrium between heating and cooling described by \refeq{att}. 

Using the attractor solution in \refeq{att}, therefore, we can find the phase-space diagram $T(n)$ from the algebraic equation without solving the full differential equation of \refeq{evol}. That is the approach that we take in this paper. To do so, we need to estimate the collapse timescale $t_{\rm col}(n)$ and the chemical abundances for the relevant species: free electrons and the primordial coolants $\hto$ and $\hd$ as functions of temperature and density. 

We estimate the collapse timescale by parameterizing $t_{\co}(n) = f t_{ff}(n)$. Here, $f$ is a factor that is approximated to be independent of temperature and density \citep{hirano_one_2014}. The approximation is reasonable because $f$ accounts for the factor by which the condensation to the loitering phase is extended, which is a well-defined epoch with a single characteristic timescale. In our analysis, we determine $f$ from the $\hto$ formation timescale $t_\hto$ (see \refsec{MStar} for a quantitative discussion).

\begin{table}[t]
    \centering
    \begin{tabular}{c|l|l}
    \hline\hline
        $k_{\rm H,1}$ & $\rm p +e \rightarrow H$&\cite{Abel1997}\\
        $k_{\rm H,3}$ & $\rm H + e \rightarrow H^-$& \cite{galli1998chemistry}\\
        $k_{\rm H,5}$ & $\rm H^- + H \rightarrow H_2 + e$ &rate irrelevant
        \\
        $k_{\rm D, 3}$ & $\rm D + H^+ \rightarrow D^+ + H$& \cite{Galli_2002} \\
        $k_{\rm D,4}$ & $\rm D^+ + H \rightarrow D + H^+$ &\cite{Galli_2002}\\
        $k_{\rm D,8}$ & $\rm D^+ + H_2 \rightarrow HD + H^+$ &\cite{Galli_2002}\\
        $k_{\rm D,{10}}$ & $\rm HD + H^+ \rightarrow H_2 + D^+$ & \cite{Gay_2011}\\
    \hline\hline
    \end{tabular}
    \caption{The minimal reaction network, which includes only the dominant formation pathways for $\hto$ and $\hd$, as well as $\hto$-$\hd$ interconversion, along with their sources.     \label{tab:reactions}}
\end{table}
\begin{table}[t]
    \centering
    \begin{tabular}{cll}
    \hline\hline
        Species & Initial Abundance & Source\\
        \hline
        $x_e$ & $2.5\times 10^{-4}$& {\sf{Recfast}} \citep{Seager_1999}\\
        $x_{\hto}$ & $7 \times 10^{-7}$&\cite{Hirata2006}\\ 
        $x_{\rm D}$ &$2.5 \times 10^{-5}$&\cite{Cooke2018}\\
        $x_{\rm D^+}$ & $6.3 \times 10^{-9}$& $x_{\rm D+}/x_{\rm D} \equiv x_{\rm H+}/x_{\rm H}$\\
        $x_{\rm HD}$ & $1.8\times 10^{-11}$&$x_{\rm HD}/x_{\rm D} \equiv x_{\rm H_2}/x_{\rm H}$\\
        \hline\hline
    \end{tabular}
    \caption{The initial fractional abundances and their sources. \label{tab:abundances}}
\end{table}
Following the argument in \citet{Tegmark1997}, we estimate the chemical abundances by adopting a minimal reaction network (\reftab{reactions}) and analytically solve for the abundances at a fixed temperature, instead of solving for the full coupled differential equations for the reaction rates. For example, we take the equation for the free electron abundance, $x_e=n_e/n_\rH$, where $n_\rH$ is the total number density of hydrogen (ionized and neutral),
\begin{align}
    \frac{dx_e}{dn_{\rH}} &= - k_{\rH,1} x_e^2n_\rH \frac{dt}{dn_{\rH}},
\end{align}
and approximate the right-hand side as
\begin{align}
   \frac{dx_e}{dn_{\rH}}  &= - k_{\rH,1} x_e^2 n_\rH\left(\frac{t_{\rm col,0}}{\sqrt{n_{\rH}/n_{\rH,0}}}\frac{1}{n_{\rH}}\right).
\end{align}
 
 Similarly for $\hto$ and $\hd$ we have: 
 \begin{align}
     \frac{dx_\hto}{dn_{\rH}} &=  k_{\rH,3} x_e n_\rH\left(\frac{t_{\rm col,0}}{\sqrt{n_{\rH}/n_{\rH,0}}}\frac{1}{n_{\rH}}\right)\\
       \frac{dx_\hd}{dn_{\rH}} &= 
     \left(k_{\hd,eff} x_{\rm D} x_e n_{\rH}  - k_{\rm D,10} x_{\rm \hd} x_{e} n_{\rH}\right)\left(\frac{t_{\rm col,0}}{\sqrt{n_{\rH}/n_{\rH,0}}}\frac{1}{n_{\rH}}\right), \label{eq:hdrate}
 \end{align}
where we have defined the effective $\hd$ formation rate
\begin{align}
    k_{\hd\rm, eff}&=k_{\rm D,3} \frac{k_{\rm D,8}}{k_{\rm D,8} + k_{\rm D,4}/x_{\rm H_2}}.
    \label{eq:khdeff}
\end{align}
In these equations, we have made several assumptions. First, that neutral $\rm D$ and $\rm H$ are not depleted and that $x_{e} \equiv x_p$ is reduced only by the recombination of neutral hydrogen. Second, that the only important channel for $\hto$ production is through $\rm H^-$. Third, that $x_{\hto}$ can be treated as constant in Eq.~\ref{eq:khdeff}, because the $\hto$ fraction is typically nearing its asymptotic value before $\hd$ production becomes significant. Finally, to derive the effective $\hd$ production rate, Eq.~\ref{eq:khdeff}, we have assumed that the $\rm D^+$ produced by reaction $k_{\rm D,3}$ is promptly converted either back into $\rm D$ or into $\hd$.
 
 As a result, we obtain the following solutions for abundances given their initial values (with the subscript 0):
\begin{widetext}
\begin{align}
    x_{\re}(n_{\rH}) &= \frac{x_{\re,0}}{1 + 2k_{\rm H,1} t_{\co,0} x_{\re,0} (\sqrt{n_{\rH} n_{\rH,0}} - n_{\rH,0})} \label{eq:xe}\\
    x_{\rm H_2}(n_{\rH}) &= x_{\hto,0} \nonumber\\
    &\quad+ \frac{k_{\rm H, 3}}{k_{\rm H,1}} \log{\left[1+ 2 k_{\rm H,1} t_{\co,0}x_{\re,0} (\sqrt{n_{\rH}n_{\rH,0}}-n_{\rH,0})\right]} \label{eq:h2}\\
    x_{\hd}(n_{\rH}) &=x_{\hd,0} \left(\frac{x_\re(n_{\rH})}{x_{\re,0}}\right)^{k_{\rm D, 10}/k_{\rm H,1}}
    \nonumber\\
    &\quad+ \frac{k_{\rm \hd, eff}}{k_{\rm D, 10}}x_{\rm D,0}\left(1- \left(\frac{x_\re(n_{\rH})}{x_{\re,0}}\right)^{k_{\rm D,10}/k_{\rm H,1}}  \right).   \label{eq:hd}
\end{align}
\end{widetext}
 \reftab{abundances} summarizes the initial abundances that we use for the computation. The rate coefficients $k_X$ are drawn from the ``primordial'' network of KROME \citep{grassi_krome_2014}, except for $k_{\rH 3}$, which is sourced from \cite{galli1998chemistry}.

Finally, to compute the cooling rate ${\cal C}$ for given abundances, we use the $\hto$ cooling rates of \citet{Hollenbach1979} and the $\hd$ cooling rate of \citet{Lipovka2005}. More recent calculations of the $\hto$ cooling rates (i.e.,~\citealt{Coppola2019}) have more complicated functional forms and differ by less than a factor of two from the rates of \citet{Hollenbach1979} in the relevant temperature regime, which is acceptable in the context of our approximate model.

We solve \refeq{att} for a halo with mass $M_{\rm Halo}$ at redshift $z$. First, we use the spherical collapse \citep[section 5.1 of][]{PBSreview} value  of the virial density of the halo $\rho_V = 178 \bar\rho_{\rm m}(z)$, and estimate the virial temperature by
\begin{equation}
    T_V = \left(\frac{4\pi}{3}\rho_V\right)^{1/3}\frac{G \mu m_H}{k_B} M_{\rm Halo}^{2/3}.
    \label{eq:TV}
\end{equation} 
We start the computation by setting the cosmological abundances (shown in \reftab{abundances}) as initial values, at $\rho_i = \rho_V/3$ (that is, somewhat before the virialization) to allow for any changes in the chemical compositions prior to virialization. Because the cooling is inefficient ($t_{\cc} > t_{\co}$) under these initial conditions, we initially assume adiabatic heating $T \propto n^{2/3}$, and correspondingly initialize the temperature at $T_i = T_V(1/3)^{2/3}$. The density and temperature then evolve along the adiabatic track until the density $n$ reaches the attractor track: $t_{\co}(n) = (\gamma-1) t_{\cc}({\vec n},T)$. Computing abundances using \refeqs{xe}{hd} requires a temperature, which during this heating phase is supplied as the geometric mean of the initial and current temperature, $\bar{T}=\sqrt{TT_i}$.

From this intercept with the $t_{\co} = (\gamma-1)t_{\cc}$ curve onward, the trajectory is determined by solving $t_{\co}= (\gamma-1)t_{\cc}$ for the temperature. 
\begin{figure}
    \centering
    \includegraphics[width=\columnwidth]{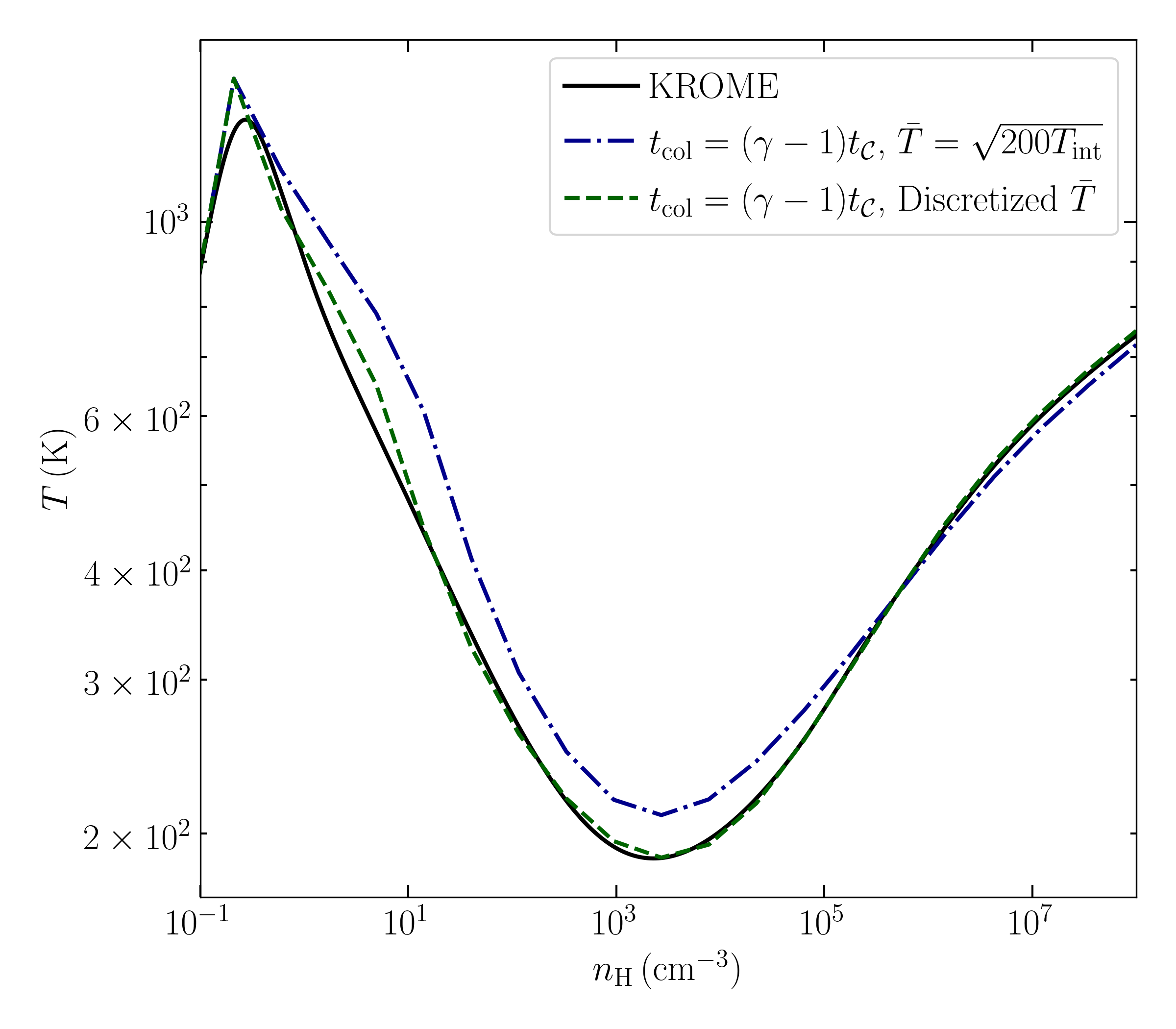}
    \caption{Evolution of the one-zone parcel in phase space for the $t_{\co} = t_{ff}$ case. The solid black line shows the result of solving the chemical network using KROME, and the other two lines show two analytic approximations. For the blue dot-dashed line, the reaction rates in Eqs.~(\ref{eq:xe}-\ref{eq:hd}) are always evaluated at a single temperature $\bar T = \sqrt{200T_{\rm int}}$ when Eq.~(\ref{eq:att}) is solved at any density. For the green dashed line, we solve Eq.~(\ref{eq:att}) at twenty logarithmically spaced steps of density $n_{\rH}\in [n_{\rm int},10^{8}\ \rm cm^{-3}]$, where to solve for the temperature $T_{n}$ at a step $n$, the reaction rates are evaluated at a temperature $\bar T = \sqrt{T_{n_{\rm prev}} T_n}$ given the temperature $T_{n_{\rm prev}}$ from the previous step $n_{\rm prev}$. The initial density and temperature are those of a halo with $M_{\rm Halo}=5\times 10^5 M_\odot$ at $z=15$.}
    \label{fig:h2onez}
\end{figure}
In the case where $f=1$ (that is, $t_{\co} = t_{ff}$), collapse occurs too quickly for $\hd$ to contribute to the cooling, and only the first two reactions in \reftab{reactions} are necessary. Moreover, the $\hto$ production rate depends weakly enough on temperature that evaluating the temperature at each density as $\bar T = \sqrt{200T_{\rm int}}$, where $\sim 200 \, \rm K$ is the molecular cooling limit and $T_{\rm int}$ is the temperature when for the first time $t_{\co} = (\gamma-1)t_{\cc}$,  can reproduce the one-zone results calculated using KROME \citep{grassi_krome_2014}, as shown in Fig.~\ref{fig:h2onez}.

\begin{figure}
    \centering
    \includegraphics[width=\columnwidth]{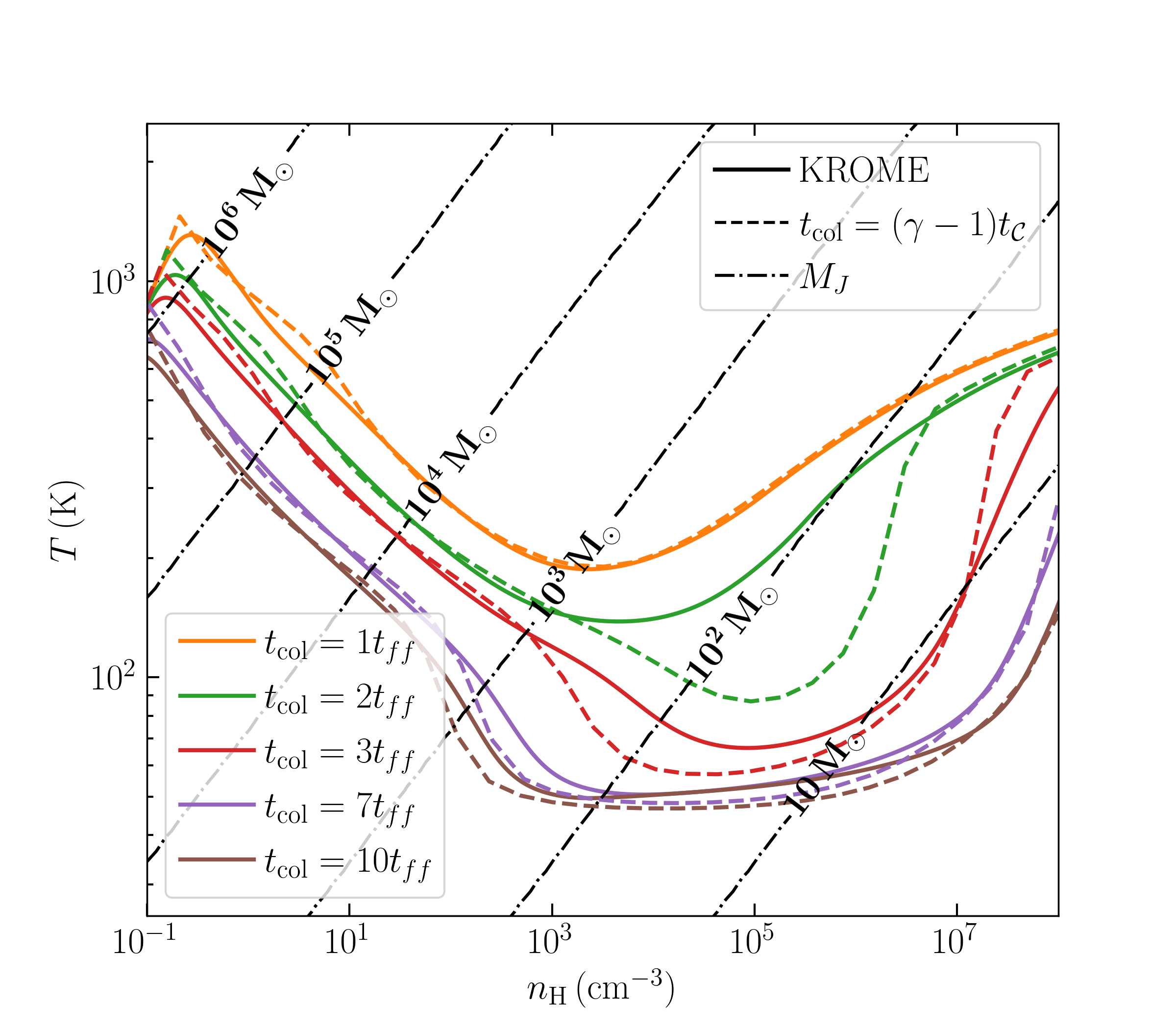}
    \vspace{-20pt}
    \caption{Evolution of the one-zone parcel in phase space for five $f\equiv t_{\co}/t_{ff}$ values. The solid lines show the results of integrating the chemical network using KROME, and the dashed lines show the solution of the algebraic relationship $t_{\cc}=(\gamma - 1) t_{\co}$, with 20 discretized density values (see the main text). The algebraic solution matches well with the KROME result. Also shown are lines of constant Jeans mass $M_J$: The $\hd$ formation lowers the minimum temperature, which reduces the Jeans mass at the critical density $n_{\rm crit}\sim10^3\,{\rm cm}^{-3}$. The initial density and temperature are those of a halo with $M_{\rm Halo}=5\times 10^5 M_\odot$ at $z=15$.}
    \label{fig:krometeq}
\end{figure}
However, for $f = t_{\co}/t_{ff}\gtrsim 3$, $\hd$ production becomes important. The relevant reaction rates, for example, $k_{\rm D,10}$, are strongly temperature dependent. To capture the temperature-dependent interaction rates, we discretize the attractor curve to 20 pieces up to a density of $n_{\rH}=10^8\,{\rm cm}^{-3}$. That is, we solve Eq.~(\ref{eq:att}) on 20 logarithmically spaced steps of $n_{\rH}$ beginning at $T_{\rm int}$ and $\vec n_{\rm int}$. In each subsequent step $n_{\rH}$, to solve for the temperature $T_{n}$, the abundances with the subscript 0 in Eqs.~(\ref{eq:xe}), (\ref{eq:h2}) and (\ref{eq:hd}) are supplied by the previous step, and the reaction rates are evaluated at a temperature $\bar T = \sqrt{T_{n_{\rm prev}} T_n}$ given the temperature $T_{n_{\rm prev}}$ from the previous step $n_{\rm prev}$. \reffig{krometeq} demonstrates an excellent match between the resulting phase-space diagram over a range of values of $f\in[1,10]$. Here, the dashed lines are the result from the algebraic approximation in this paper, and solid lines are the solution of the full ODE using the KROME \citep{grassi_krome_2014} package. Note that although the results disagree somewhat for $f=2$ due to our omission of the subdominant $\hd$ dissociation reaction $\hd + \rm H \rightarrow  \hto + \rm D$, our model will require only the temperature at the critical density $n_{\rm crit} \sim 10^3 \, \rm cm^{-3},$ (\refsec{mcloud}) at which point the curves still agree very closely. With \refeq{att} we have transposed the problem of solving a system of coupled ordinary differential equations (already a great simplification of the full, three dimensional problem) to a root finding problem involving inexpensive function calls at a small number of grid points.
\section{Stellar Mass}
\label{sec:MStar}

We are now in a position to estimate the final stellar mass. 
Here for simplicity, we assume that only a single protostar forms at the center and grows by accretion via a disk until stellar radiation evaporates the cloud.
We will define the collapse timescale $t_\co$ (that is, fixing $f$) to estimate the mass of the collapsing gas cloud. We then determine the final stellar mass in two steps. First, we compute the mass of the initial stellar core that forms before Kelvin-Helmholtz contraction dominates over accretion. We then compute the mass accreted onto this stellar core: the accretion rate is given by dividing the cloud mass with the viscous timescale, and the accretion shut-off time is defined as when the star-forming cloud is completely ionized by the radiation from the protostar. Finally, we compare our result with the fitting formula for the Pop. III star mass derived from 2D radiative hydrodynamic simulations in \citet{hirano_one_2014}, which also assume a single protostar in each star-forming disk.

Actually, recent hydrodynamic simulations of primordial star formation have converged on the picture that Pop.~III star-forming disks typically undergo fragmentation and produce multiple protostars \citep{Turk2009,Stacy2010,Clark2011,Greif2011,susa_merge_2019,Liu2020,Haemmerl__2020,klessen2023stars} unless stabilized by strong magnetic fields \citep[e.g.,][]{Sharda_2020,Sharda_2021,Hirano2022}. The formation and evolution of protostars during disk fragmentation (e.g., growth by competing accretion, mergers, and ejections by gravitational scatters) are still poorly understood due to the limited resolution and/or time coverage of current simulations. It is likely that multiple protostars will survive and grow to comparable masses by the moment of cloud evaporation \citep{Sugimura2020,Sugimura2023,Park2023,Park2023a}. In this case, the strength of ionization feedback will be overestimated by the assumption that the star-forming disk only feeds one central protostar, since the efficiency of producing ionizing photons increases with stellar mass. Therefore, the final stellar mass estimated by our idealized model should be regarded as a lower limit. We defer a more general analysis that includes the situation of multiple protostars to future work.

\subsection{Collapse Timescale and Mass of the Cloud}
\label{sec:mcloud}
As discussed in Sec.~\ref{sec:Timescales}, the collapse timescale is determined by the $\hto$ production timescale. If the collapse timescale is long, significant $\hd$ can form leading to a lower minimum temperature \citep{Ripamonti_2007}. It is this minimum temperature that ultimately sets the mass of the Jeans-unstable cloud. Relatively massive halos form $\hto$ efficiently within their initial (virial) free-fall timescale. On the other hand, low-mass halos ($M_{\rm halo}\sim10^5\,M_\odot$ at $z\sim20$, for example) are initially too cold to rapidly form $\hto$ and cool further. 

To be specific, we define the collapse timescale as 
\begin{equation}
    t_{\co} \approx f t_{ff},
\end{equation}
with
\begin{equation}
   f= \max{(t_{\hto}/t_{ff},1)}
\end{equation}
where $t_\hto$ is the $\hto$ production timescale that we defined in \refsec{Timescales}. The timescales $t_\hto$ and $t_{ff}$ are evaluated at the virial density and temperature to determine the value of $f$, which is assumed to be constant during collapse \citep{hirano_one_2014}. Quantitatively, $t_\hto$ is the timescale required to produce enough $\hto$ to satisfy $t_{\cc}=t_{ff}$ at the virial density and temperature. That is, the critical $\hto$ abundance for cooling $x_{\hto, \cc}$ is defined by
\begin{equation}
    t_{\cc}(n_V, T_V,x_{\hto, \cc})-\frac{{\cal E}_{th}(T_{\hto})}{{\cal C}(n_V,T_V, x_{\hto, \cc})} = t_{ff}(n_V),
\end{equation}
where the subscript $V$ stands for quantities in the initial, virial equilibrium and 
 $T_{\hto} \approx 200~{\rm K}$  is the minimum temperature achievable by $\hto$ cooling at the loitering phase. The second term on the left-hand side signifies that we are interested in the time until arrival at the loitering phase at $T=T_\hto$. Note that formally the left hand side becomes negative for $T_V < T_{\hto}$. However, Pop.~III star formation in such low mass, low redshift halos is likely very uncommon, and we do not consider this case. Then, with the $\hto$ production rate 
\begin{equation}
    \dot x_{\hto} = k_{\rm H,3}(T_V) n_{{\rm H}, V}  x_{e,V},
\end{equation}
we have
\begin{equation}
    t_{\hto} = \frac{x_{\hto, \cc}}{\dot x_{\hto}}.
\end{equation}
Once $f$ is determined, we solve the chemical-thermal network as described in \refsec{GasChem}, with $t_{\rm col,0}=f t_{ff,V}$. In \reffig{krometeq}, we find that the $\hd$ production indeed lowers the minimum achievable temperature. 

We then extract the mass of the cloud from the $n$--$T$ trajectory as $M_c = M_J(n_{\rm crit}, T(n_{\rm crit})),$ where $n_{\rm crit} = 1.75 \times 10^3 \, \rm cm^{-3}$ characterizes the loitering phase. In a two level system, the critical density is the density at which the $\hto$ collisional de-excitation rate equals the spontaneous radiative decay rate, which remains a useful heuristic for the multi-level system. For both large and small values of $f$ (corresponding to efficient and inefficient $\hd$ formation, respectively) the critical density also approximately corresponds to the minimum temperature. However, for the transitional values of $f \sim 3$ the minimum temperature occurs at higher densities as $\hd$ continues to form later in the collapse. Even for these cases, we persist in extracting Jeans mass from the temperature at $n_{\rm crit}$ because, in order to affect the mass of the cloud, the $\hd$ must be formed \textit{before} $\hto$ cooling becomes inefficient at $n_{\rm crit}$.

Note that while \citet{hirano_one_2014} have discussed rotation as an important determinant of $t_\co$, their results reveal little correlation between the cloud mass and the rotational parameter. This is because, typically, rotation becomes an important stabilizing force only after gravitational instability sets in. An unusually high angular momentum is required for rotational support to develop before the loitering phase. Instead, we argue that rotation determines the infall rate onto the proto-star once the cloud mass is fixed.

\subsection{Accretion Rate}
The initial infall rate, as assumed in the KROME calculation [\refeq{nevol}], is
\begin{equation}
    \dot M_\star = M_c/ t_{\co}.
\end{equation}
However, the accretion rate in later stages and on small scales is typically limited by angular momentum transport. To model this, we adopt the $\alpha$-disk parameterization \citep{Shakura1973} for the viscosity $\nu = \alpha h c_s$ with $h$ being the scale height. For the thin disk, 
\begin{equation}
\frac{h}{R}\simeq \frac{c_s}{v_c} \,\to\,   \nu = \alpha c_s^2/\Omega,
\end{equation}
where $R$ is disk radius, $v_c=\Omega R$ (with $\Omega$ the angular velocity) is the circular velocity, the viscous timescale becomes
\begin{equation}
    t_\nu = \frac{R}{\nu/R}= \frac{R^2\Omega}{\alpha c_s^2}.
\end{equation}
We now make the assumption that the specific angular momentum, $J \propto R^2\Omega$ is conserved, which is reasonable because in a hot, thick Pop III star-forming disk in the absence of strong (regular) magnetic fields, there are no processes (e.g., resonant torques and outflows) that can efficiently transport angular momentum away. We can thus evaluate the viscous timescale of the disk at any time using the value of $R^{2}\Omega$ at the critical density $n_{\rm crit}$ as
\begin{equation}
  t_\nu =\frac{R_{\rm crit}^{2}\Omega_{\rm crit}}{\alpha c_{s}^{2}}= \frac{\sqrt{3 \beta G M_c c_{s,\rm crit}t_{ff, \rm crit}}}{\alpha c_s^2},
\end{equation}
 with the subscript ``${\rm crit}$'' indicating the value of the quantity evaluated at the critical density. Here, we have assumed that at $n_{\rm crit}$ the radius of the cloud is of order the Jeans length, $R_{\rm crit}\approx c_{s, \rm crit}t_{ff, \rm crit}$, and introduced the spin parameter 
\begin{equation}
    \beta \equiv \frac{v_c^2}{3|\Phi|} = \frac{\Omega_{\rm crit}^2R_{\rm crit}^3}{3GM_c}
\end{equation}
as the ratio of rotational to gravitational energy in the cloud at $n_{\rm crit}$ to characterize the degree of rotation.
 
 Initially, $t_{\co} > t_\nu$. As the density increases, $t_{\co}$ decreases as $1/\sqrt n$ while $t_\nu$ varies only due to the factor-of-few change in sound speed as the gas cools and then heats. Hence, eventually $t_\nu > t_{\rm col}$. Therefore, the accretion rate onto the protostar is ultimately set by the viscous timescale. We estimate the accretion rate as 
\begin{equation}
    \dot M_\star = M_c/t_\nu.
    \label{eq:mdotstar}
\end{equation}

In calculating $t_\nu$ we assume $T=1000 \, \rm K$, typical of the molecular disk. As the fiducial case we adopt $\beta = 0.3$, which is a typical value in \cite{hirano_one_2014}. Here, we take $\alpha = 1$. The characteristic values of $\alpha$ in \cite{hirano_one_2014} are a factor of a few lower, but we find that applying $\alpha=1$ in Eq.~(\ref{eq:mdotstar}) better matches the accretion rates in that work (see App.~\ref{app:comp}). We assume $\dot M_\star$ is constant, which is equivalent to replacing $\dot M_\star(t)$ with its average value. Note that \cite{Liu2020} gives a semi-analytic universal solution for Pop.~III growth $M \propto t^{4-3 \gamma_{eff}} \approx t^{0.7}$ (with $\gamma_{eff} = 1.09$ as the effective polytropic index of gas in primordial star-forming disks \citep{Omukai_1998}) which is not far from the constant growth rate.

\subsection{Mass, Radius, and Luminosity}
Once the proto-stellar core is formed, the radiation field from the core competes against the accretion flow to determine the evolution during the protostar phase. 

For Pop. III stars, \citet{Stahler1986} have studied the evolution of protostars while accretion dominates the dynamics. We adopt their analytical estimate for the protostar core radius: 
\begin{equation}
    R_\star = 26 R_\odot \left(\frac{M_\star}{M_\odot}\right)^{0.27}\left(\frac{\dot M_\star}{10^{-3}\, \rm M_\odot\, yr^{-1}}\right)^{0.41},
    \label{eq:rstar}
\end{equation}
which is surrounded by optically thick radiative precursor with a photospheric radius of $R_{\rm ph}=1.4 R_\star$. Here, the star symbol denotes protostellar quantities. Also, when the opacity is dominated by electron scattering, the luminosity of the hydrostatic equilibrium object is proportional to $M_\star^3$, and \citet{hosokawa_rapidly_2012} find the following approximate relationship:
\begin{equation}
	L_\star \simeq 10 L_\odot\left(\frac{M_\star}{M_\odot}\right)^3\,.
\end{equation}

Two timescales are relevant in determining the contracting mass scale of the protostellar core: the accretion timescale 
\begin{equation}
    t_{\rm acc} = \frac{M_\star}{\dot M_\star},
\end{equation}
and the Kelvin-Helmholtz timescale,
\begin{equation}
    t_{\rm KH} = \frac{G M_\star^2}{R_\star L_\star}.
\end{equation}
Initially, the accretion timescale is short compared to the Kelvin-Helmholtz timescale, and the protostar expands according to \refeq{rstar}. Eventually, however, the mass and luminosity growth of the protostar cause Kelvin-Helmholtz contraction to dominate over accretion: $t_{KH} < t_{\rm acc}$, which happens at $M_{\rm eq}$ \citep{hosokawa_rapidly_2012}:
\begin{equation}
    M_{\rm eq} \simeq 15\, {\rm M_\odot} \left(\frac{\dot M_\star}{10^{-2} \rm M_\odot yr^{-1}}\right)^{0.26}\,,
\end{equation}
at which point the protostellar cores with luminosity less than the Eddington luminosity, $L(M_{\rm eq}) < L_{\rm Edd}$, begin to contract. The following energy balance equation can model the contraction:
\begin{equation}
\frac{d}{dt}\left(\frac{W}{2}\right)
=
\frac{3}{2(5-n)}\frac{d}{dt}\left(\frac{GM_\star^2}{R}\right)
= L(M_\star),
\label{eq:lr}
\end{equation}
where $W$ is the gravitational binding energy of the protostar and we assume that each step of contraction plus accretion maintains a new virial equilibrium by radiating the energy difference. Here, $n$ is the polytropic index $P\propto\rho^{(n+1)/n}$ and we choose $n=3$, consistent with the Eddington-beta model \citep{Eddington1926}, which is reasonably accurate for massive stars.

Again, assuming the hydrostatic equilibrium protostellar core with constant opacity, dominated by electron scattering, $L_\star\propto M_\star^3$, we solve the energy balance equation (Eq.~(\ref{eq:lr})) to find the radius-mass relationship:
\begin{equation}
    R(M_\star)
	= R_{\rm eq} 
	\left[\frac{(M_\star/M_{\rm eq})^2}{((M_\star/M_{\rm eq})^4 - 1)/3 + 1}\right]\,
    \label{eq:rstar2}.
\end{equation}

Eventually, the luminosity of the collapsing proto-star reaches the Eddington limit \citep{hosokawa_rapidly_2012},
 \begin{equation}
     L_{\rm Edd} = 3.8 \times 10^6 \left(\frac{M_\star}{100 M_\odot}\right) L_\odot\,,
 \end{equation}
from which point radiation pressure prevents the contraction of the envelope \citep{hosokawa_rapidly_2012, hirano_one_2014}. Then, the protostellar radius must either oscillate or grow. Substituting the Eddington luminosity into \refeq{lr} gives a nearly constant radius. However, \cite{hosokawa_rapidly_2012} find that once the protostars reach the Eddington luminosity, the gravothermal evolution contracts the core while expanding the envelope. The result of these more complicated dynamics is to grow the characteristic radius as $R_\star \propto M_\star^{0.5}$, which we adopt here. Note that \cite{hosokawa_rapidly_2012} find that for very high accretion rates $\gtrsim 0.1 \, \rm M_\odot/yr$ the proto-star begins to expand even before $L_{\rm Edd}$, an effect which we neglect, both because the underlying physics are complex and because this mechanism principally operates for rotation rates lower than our fiducial $\beta = 0.3$. This omission will lead to an underestimate of the masses of the most massive, slowly rotating stars.
 \begin{figure*}[h!t]
     \centering
     \includegraphics[width=0.95\textwidth]{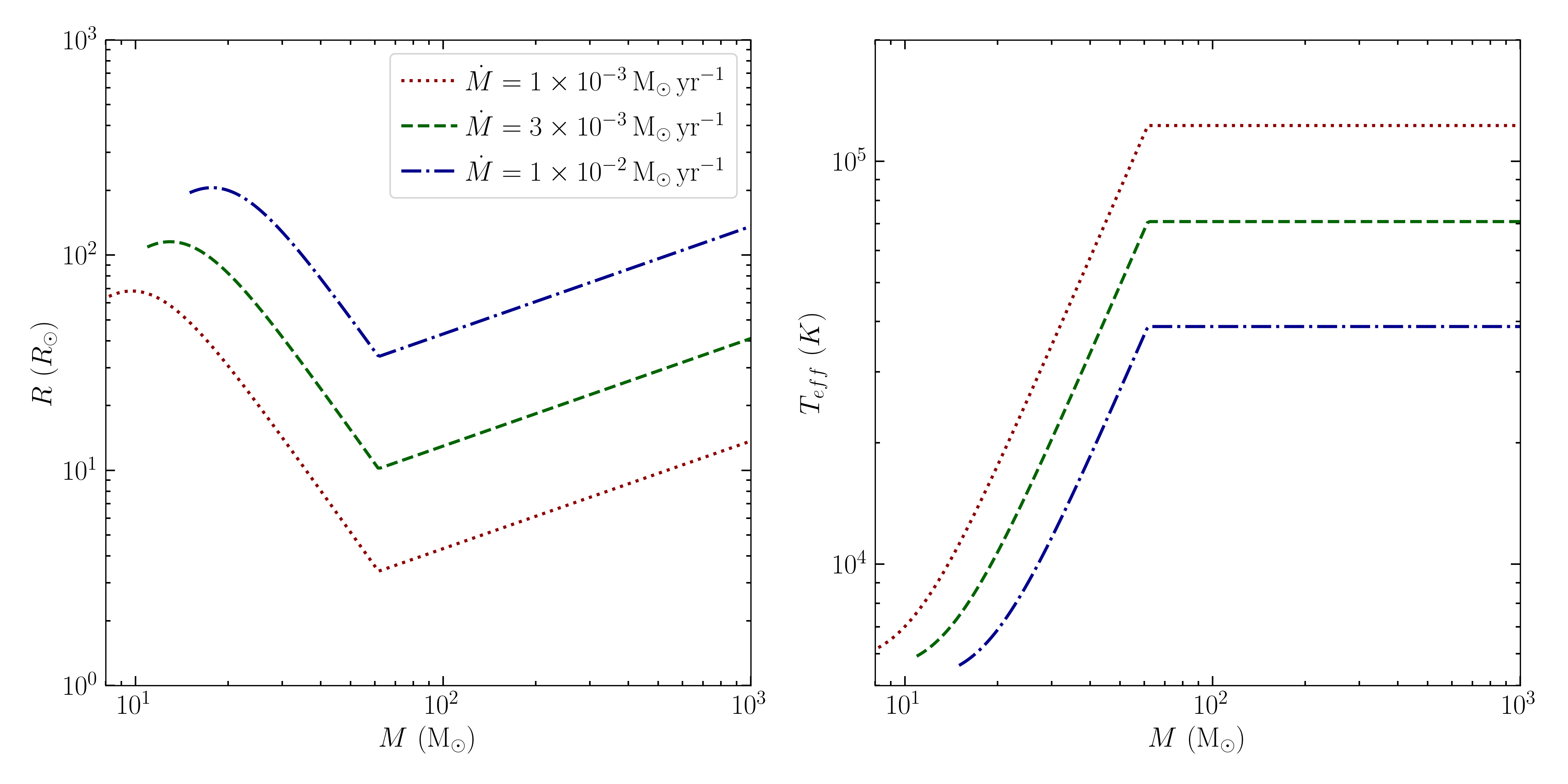}
     \vspace{-10pt}
     \caption{The radius and effective temperature implied by \refeq{rstar} (radius at $t_{eq}$), \refeq{rstar2} (evolution to the Eddington luminosity), and $R_\star \propto M_\star^{0.5}$ (upon reaching the Eddington limit). Compare with Fig.~12 of \cite{hirano_one_2014} and Fig.~5 of \cite{hosokawa_rapidly_2012}.}
     \label{fig:rt}
 \end{figure*}
 
 We show the evolution of radius and effective temperature of the protostars in \reffig{rt} for three different accretion rates.

 \subsection{Ionizing Feedback and Stellar Mass}
 Equipped with the effective temperature and radius that we estimated in the previous section (see \reffig{rt} for the result), we can also estimate the ionizing photon flux, $S_{\rm EUV}$, by integrating the black body spectrum. 
 
 The interaction between the UV flux and the hydrodynamical gas launches a shock. We assume that the shock is launched into the cloud at $t_{\rm eq}$ (i.e. the time when the mass reaches $M_{\rm eq}$)  which is the characteristic epoch where the protostar begins to emit significant UV radiation. Once the shock is launched, it homogenizes the downstream medium and reduces its density. Initially, the UV radiation is trapped behind the shock front by recombinations. Eventually, the density behind the shock (that is, towards the center of the cloud) falls low enough that recombination is no longer efficient. Then, the ionization front escapes the shock front, rapidly ionizing the cloud and shutting down accretion. This cloud breakout time $t_B$ is estimated in \citet{Alvarez_2006} using the similarity solution of \citet{Shu_2002} for a singular isothermal sphere (SIS) with a density profile of $\rho\propto r^{-2}$ as the initial condition. There, $t_B$ is calculated by equating the ionizing photon flux at $t_{B}$ after the shock onset to the recombination rate behind the shock front:
 \begin{equation}
     S_{\rm EUV}(t_B) = 4 \pi \alpha_B\int_0^{r_{sh}(t_B)} dr r^2 n(r,t_B)^2, \label{eq:seuv}
 \end{equation}
 where $\alpha_B$ is the case-B recombination rate coefficient for hydrogen at the temperature characteristic of the photo-heated gas $\sim 10^4 \, \rm K$, and $r_{sh} = x_s c_s t$  (with $c_s$ the sound speed in the shocked gas) is the radius of the shock front, and the normalization of the density profile $n$ is proportional to the temperature $T_{\rm SIS}$ of the initial SIS, which we take to be the temperature at the loitering phase, i.e. $T_{\rm SIS}=T(n_{\rm crit})$. When the shocked gas is much hotter than the surrounding isothermal sphere (which holds in all cases we encounter), we have $x_s \approx 2.56$. 

 \cite{Alvarez_2006} provide the following approximate relationship for the breakout time, which we find matches the exact result to within $\sim 10 \%$:
 \begin{equation}
     \begin{split}
    t_B &= 6.5 \times 10^4 {\, \rm yr}\left(\frac{c_s x_s}{40 \, \rm km \, s^{-1}}\right)^{-1} \left(\frac{T_{\rm SIS}}{300 \, \rm K}\right)^2\\
    &\times\left[\frac{S_{\rm EUV}(t_B)}{3\times 10^{50} \, \rm s^{-1}}\right]^{-1}.     
 \end{split}
 \label{eq:tb}
 \end{equation}
Note that helium at the cosmological mass fraction $Y = 0.24$ enters the calculation of the sound speed via the mean molecular weight. We assume the first ionization of helium is coupled with that of hydrogen, and correspondingly have multiplied the prefactor in Eq.~\ref{eq:tb} by $x_{\rm H}^{-2}=1.16$ compared to \cite{Alvarez_2006} (who neglected the consumption of ionizing photons by helium) given the primordial number fraction of hydrogen nuclei $x_{\rm H}=0.927$. We finally solve Eq.~(\ref{eq:tb}) for $t_B$, and then compute the final stellar mass as 
\begin{equation}
    M_\star = M_{\rm eq} + \dot M_\star t_{B}.\label{eq:m_star}
\end{equation}
The solution in \citet{Shu_2002} ignores the gravity from the central star and the fact that the density profile in the central region is shallower than $\rho\propto r^{-2}$ in the presence of the star-forming disk. These effects tend to slow down the propagation of the ionization front in the early stage so that the ionized region can be trapped in the center for an extended period $t_{\rm trap}\gtrsim 20$~kyr before breaking into the cloud \citep[][]{Jaura2022}. Since this initial trapped phase is not considered in our model, the final stellar mass can be underestimated\footnote{Assuming that the \citet{Shu_2002} solution describes the dynamics of the ionized gas well after the shock breaks into the cloud, our model can be easily extended to take into account the initial trapped phase by replacing $t_{B}$ with $t_{B}+t_{\rm trap}$ in the right-hand sides of Eqs.~(\ref{eq:seuv}-\ref{eq:m_star}). However, the dependence of the trapping timescale $t_{\rm trap}$ on cloud/halo properties is still unclear. We therefore defer such extension to future work. In this paper, we aim to compare the results of our simple model with those from the 2D simulations in \citet{hirano_one_2014}, where the central disk scale height as well as the initial trapped phase is unresolved.}.

\begin{figure}
    \centering
    \includegraphics[width=\columnwidth]{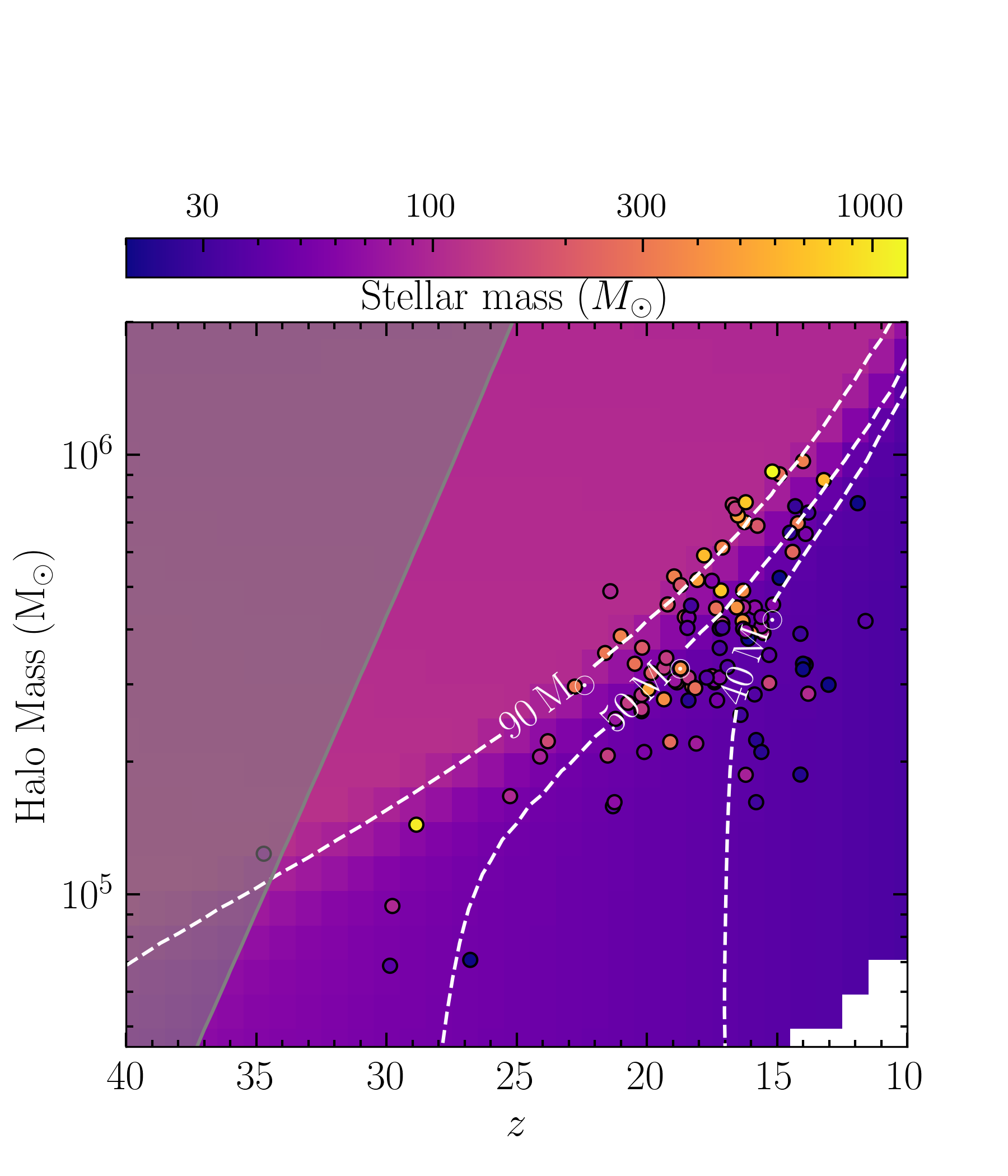}
    \vspace{-20pt}
    \caption{The predicted stellar mass in our model (background color, contours)  compared with the primordial stars found in the simulations of \cite{hirano_one_2014} (colored circles). The gray shaded region indicates halos which will not typically appear in the simulation volume of \cite{hirano_one_2014} in the Press-Schechter formalism. In the bottom right, $T_V < T_{\hto}$ and we do not calculate a stellar mass. Our model is accurate to within a factor of a few where \cite{hirano_one_2014} have robust data (between $\sim 30$ and $\sim 300$ solar mass).}
    \label{fig:mcomp}
\end{figure}

Having relied on the timescales and analytical estimate, our estimation of the final stellar mass only takes a few seconds. On the other hand, with a series of 2D radiative hydrodynamic simulations of protostar accretion from initial conditions of primordial star-forming clouds produced by cosmological hydrodynamic (zoom-in) simulations, \citet{hirano_one_2014} developed a sample of $\sim 100$ Population III stars. 
\reffig{mcomp} shows that our estimation very closely resembles the result of the state of the art simulations! The figure shows the predictions of our model for $\beta = 0.3$ as a function of halo mass and redshift. We have also plotted the stellar sample of \cite{hirano_one_2014} (dots) and the line corresponding to a halo abundance of one in the simulation volume of \citet{hirano_one_2014} in the Press-Schechter Formalism \citep{Press1974, Sheth2001}. Halos become common enough to typically appear in the simulation volume of \cite{hirano_one_2014} only to the right of this line. However, \cite{hirano_one_2014} employ additional selection criteria to ensure that each star-forming cloud is pristine. In the bottom right corner of the figure, $T_V < T_\hto,$ and our calculation of $f$ does not apply. As suggested above, primordial star formation in this parameter space is not important.

\begin{figure}
    \centering
    \includegraphics[width=\columnwidth]{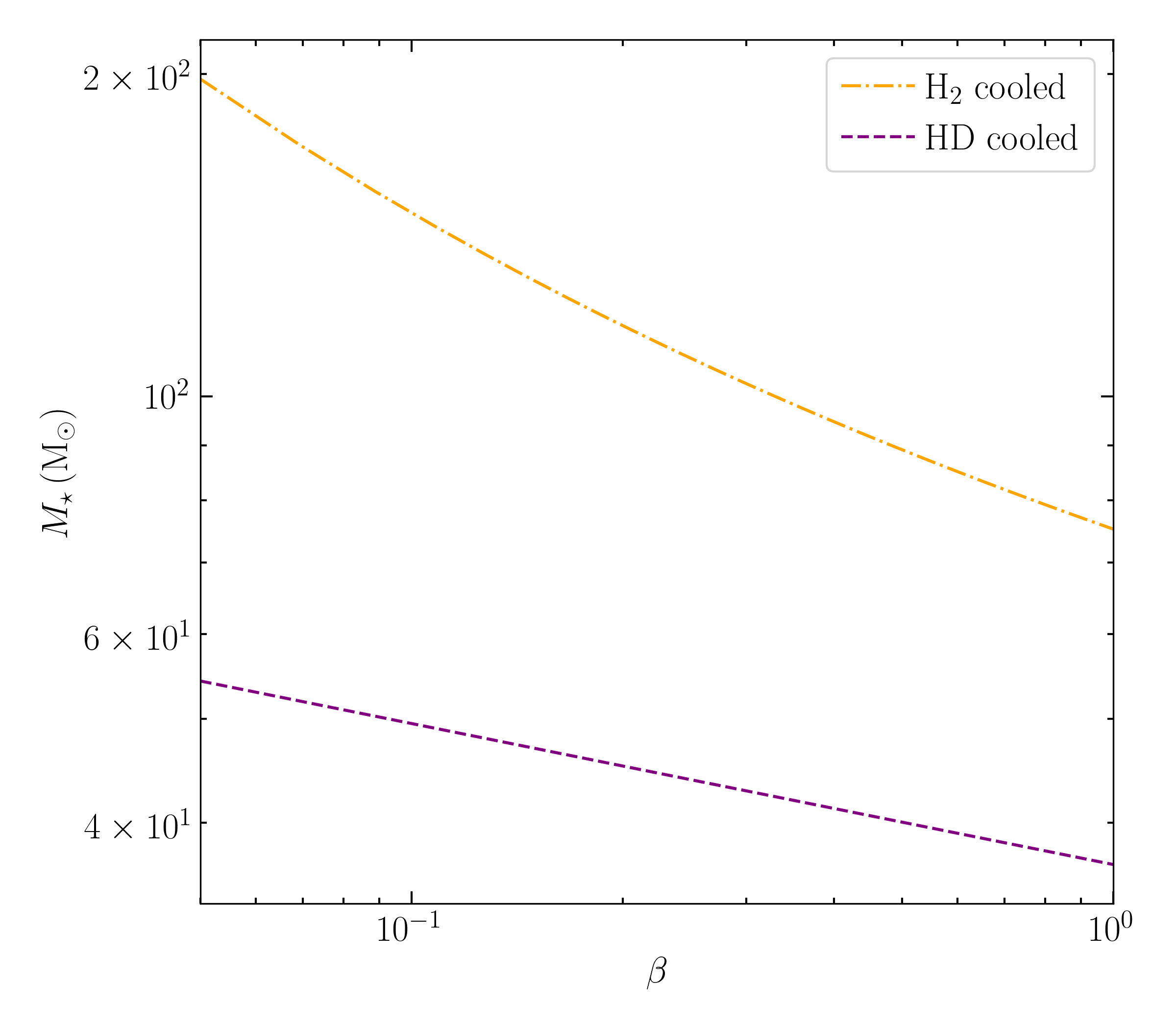}
    \vspace{-20pt}
    \caption{The role of the rotation parameter $\beta$ in the final stellar mass for an $\hto$ cooled ($10^6 \, \rm M_\odot$, gold) cloud and $\hd$ cooled ($10^5 \, \rm M_\odot$, purple) cloud, both at $z=25$, showing the stronger dependence on $\beta$ at higher cloud masses. }
    \label{fig:beta}
\end{figure}

Our model succeeds quite well in predicting the transition between $\hto$ cooled clouds (final stellar mass of hundreds of solar masses) and $\hd$ cooled clouds (final stellar masses of order tens of solar masses). Over the range where \cite{hirano_one_2014} have reasonable statistics ($\sim 30-300 \, \rm M_\odot$), our model agrees with the simulation results to within a factor of a few. However, our model fails to produce the most massive stars observed in \cite{hirano_one_2014}, a fact which cannot be entirely explained by our choice of fixed $\beta =0.3$ in Fig.~\ref{fig:mcomp} (see Fig.~\ref{fig:beta}). There are several physically reasonable explanations which may contribute to this discrepancy, which are explored in more detail in App.~\ref{app:comp}. First, the rapidly accreting stars of \cite{hirano_one_2014} can grow by around $100 \, \rm M_\odot$ after HII breakout due to the geometry of the accretion disk which is not captured by our spherically-symmetric breakout model. Relatedly, our assumption of a thin disk in calculating $t_\nu$ breaks down for the slowly rotating clouds which produce the most massive stars, leading to an underestimate of the accretion rate.  Finally, as mentioned above our model fails to account for the finding of \cite{hosokawa_rapidly_2012} that for very high accretion rates the proto-star never undergoes a period of contraction. The low surface temperature associated with this large stellar radius is necessary to attain stellar masses $\sim 1000 \, M_\odot$. These facts can largely explain the systematically lower masses predicted by our model at high accretion rates.

We also show in Fig.~\ref{fig:beta} the effect of varying $\beta$ on a low mass, $\hd$ cooled cloud and a high mass, $\hto$ cooled cloud at redshift 25. In our model, the largest stellar mass attainable with a minimal realistic rotation parameter $\beta \approx 0.05$ is $M\sim 200 \, \rm M_\odot$, with an accretion rate of $\sim 0.01 \, \rm M_\odot/yr$. Rotation has a larger effect at higher accretion rates (lower $\beta$), because the efficiency of the UV feedback has a strong dependence on the temperature at which the surface of the Eddington radiating stars is ``frozen'' (Fig.~\ref{fig:rt}). 
\section{Conclusion \& Discussion}
\label{sec:Discussion}
We have developed a simplistic model of the formation of Pop.~III stars in the center of collapsing primordial gas clouds. The model consists of two parts. The first determines the chemical-thermal evolution of the collapsing gas cloud using the dynamical-thermal equilibrium relation $t_{\co} = (\gamma-1)t_{\cc}$. The second relates the mass and spin parameter of the collapsing cloud to the final mass of the star. For a typical value of the spin parameter, this model agrees with the masses predicted by sophisticated simulations \citep{hirano_one_2014} to within a factor of a few for stellar masses between $\sim 30-300 \, \rm M_\odot$. However, the model struggles to produce the most massive stars seen in simulations, for which the aspherical accretion geometry strongly affects the final mass (see App.~\ref{app:comp}). Moreover, simulations such as those of \cite{hirano_one_2014} which do not resolve the trapped phase of HII region \citep{Jaura2022} and produce only one star per cloud may underestimate the star formation efficiency. Since our model also assumes one star per cloud and ignores the trapped phase, the predicted final stellar mass is best understood as a lower bound.

The model makes liberal use of the $\approx$ symbol. Although it contains no explicit free parameters, there are implicit choices involving various order unity factors which can bring the model into better or worse agreement with simulations. These include the numerical prefactors in the sound crossing and free-fall timescales, the value of the viscosity parameter $\alpha$, the characteristic temperature of the disk used to evaluate the viscous timescale $t_\nu$, and the characteristic temperature of the singular isothermal sphere used in \refeq{seuv}. The overall trends are robust to these choices. 

We comment briefly that based only on the chemical-thermal evolution of the gas (\refsec{GasChem}) two alternative estimates of the stellar mass are already possible. First, the Jeans mass at the loitering phase can be multiplied by some global efficiency factor $M_\star = \epsilon M_c$, where to match the characteristic $\sim 100 \, \rm M_\odot$ Pop.~III stellar mass $\epsilon \sim 0.1$. However, in our model the efficiency factor depends on the cloud mass. We find that $\epsilon$ is a factor of a few larger for smaller, $\hd$ cooled clouds than for larger, $\hto$ cooled clouds. This is because while smaller clouds lead to slower accretion, the slower accretion in turn leads to a longer period of growth before the breakout of the ionization front. Alternatively, one could estimate $M_\star = N M_{l}$, where $M_l \approx 1.4 M_\odot \mu^{-9/4} \left(\frac{k_B T}{m_p c^2}\right)^{1/4}$ is the opacity limited minimum fragment mass \citep{Rees1976}, evaluated at the minimum temperature of the gas \citep{Shandera_2018, Singh_2021}, and $N \sim 10^4$ (as the effective number of fragments that merge to make the central star) to produce reasonable stellar masses. In this case, if $N$ is constant, the weak dependence of $M_l$ on $T$ means that the temperature difference between $\hto$ and $\hd$ cooled clouds amounts to less than a factor of two difference in $M_l$ and hence in $M_\star$. Additionally, neither of these simpler estimates include the role of angular momentum.

Our model correctly determines the order-of-magnitude mass scale of the first stars, and the dependence of that mass scale on redshift, halo mass, and the rotation parameter. These dependencies, which were initially emergent properties in radiation-hydrodynamic simulations, are here distilled to simple timescale arguments. If Pop.~III stars are observed, this kind of understanding will provide a powerful lever to calibrate simulations against observations. In the meanwhile, this model can be inserted in simulations of cosmological volumes at a low computational cost, allowing a novel treatment of the metal enrichment of the universe and subsequent reionization history. 

Another possible application is to the study of dissipative dark matter, which can itself cool to eventually form compact objects \citep{Shandera_2018, Hippert_2022, Gurian2022,Ryan2022}. These objects could have masses and compactnesses impossible under ordinary stellar astrophysics, leading to distinctive gravitational wave signatures. However, given the large model space there is a need for simple, inexpensive, and reasonably accurate models of the compact object formation process. The core ideas present in this model could be generalized to other dissipative physics, providing just such a tool. 

Finally, although this work only considers the formation of the massive central star in each cloud, \cite{Liu2020} found that the formation of Pop.~III star clusters by disk fragmentation can be described by simple scaling laws that capture the key trends in 3D hydrodynamic simulations of primordial star-forming clouds. Our model can be generalized and improved to consider Pop.~III star clusters and self-shielding in aspherical accretion flows using the results of 3D radiative hydrodynamic simulations that follow the growth and feedback of multiple protostars \citep[e.g.][]{Sugimura2020,Sugimura2023,Park2023a,Park2023}, which is an intriguing direction for future research.

\acknowledgements
We thank Naoki Yoshida for helpful comments and for tracking down and sharing the data from \cite{hirano_one_2014}. This work was supported at Pennsylvania State University by NASA ATP Program No. 80NSSC22K0819. DJ is also supported by KIAS Individual Grant PG088301 at Korea Institute for Advanced Study. BL is supported by the Royal Society University Research Fellowship. Research at Perimeter Institute is supported in part by the Government of Canada through the Department of Innovation, Science and Economic Development Canada and by the Province of Ontario through the Ministry of Colleges and Universities.

\bibliographystyle{aasjournal}
\bibliography{popiii.bib}

\appendix
\section{Detailed Comparison with Simulations}
\label{app:comp}
In addition to the fit to the final stellar mass [\refeq{mstarfit}], \cite{hirano_one_2014} provide fits to the accretion rate given the cloud mass and $\beta$ and to the final stellar mass given the accretion rate. These are:
\begin{align}
    \dot{M_\star} &= 2.8 \times 10^{-3} \, {\rm M_\odot \, yr^{-1}} \left(\frac{M_{c}}{350 \, M_\odot}\cdot \frac{0.3}{\beta}\right) \label{eq:mdotfit}\\
    M_\star &= 100 \, {\rm M_\odot} \left(\frac{\dot M_\star}{2.8\times 10^{-3}\, M_\odot \, yr^{-1}}\right)^{0.8}. \label{eq:mfrommdot}
\end{align}
Finally, \citet{hirano_one_2014} fit the dependence of the stellar mass on halo mass and redshift as 
\begin{equation}
    M_\star = 100 {\, \rm M_\odot\,}\left(\frac{1+z}{20}\right)^{3}\left(\frac{M_{\rm halo}}{3\times 10^5 \, \rm M_\odot}\right)^{2}.
    \label{eq:mstarfit}
\end{equation}
We can apply these estimates to better understand the discrepancies between our model and the results of \cite{hirano_one_2014}, by replacing the relevant pieces of our analytic model with these fits. This comparison will also demonstrate the flexible, modular nature of our model. The fiducial model aims for physical intuitiveness throughout the star formation process, but depending on the application it can easily be modified to provide greater fidelity to simulations. 

\begin{figure}
    \centering
    \includegraphics[width=0.5\columnwidth]{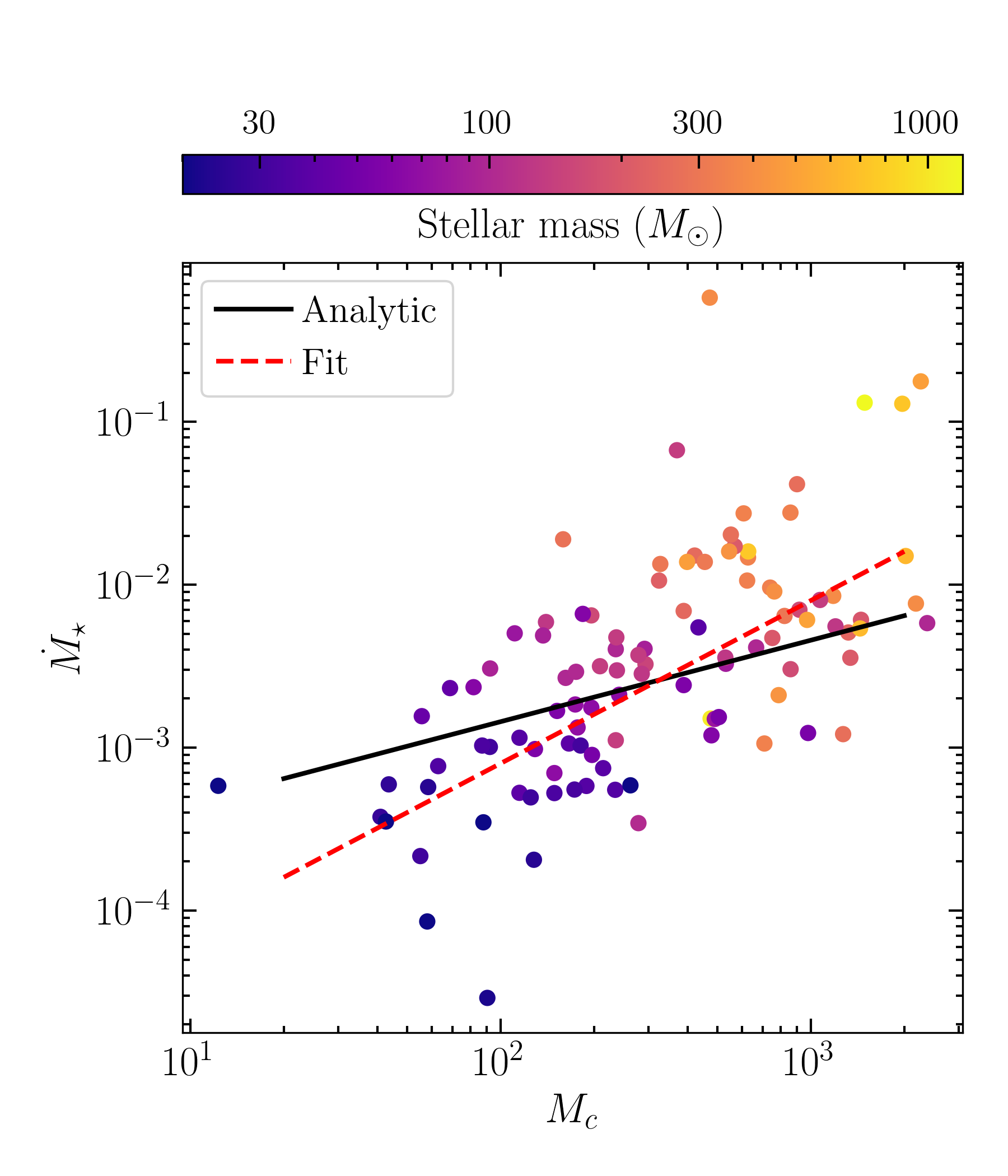}
    \vspace{-10pt}
    \caption{The accretion rate from Eq.~\ref{eq:mdotstar} (black) and from Eq.~\ref{eq:mdotfit} (red, dashed) and the sample of \cite{hirano_one_2014}, plotted against the mass of the star-forming cloud. Although the analytically calculated accretion rate is broadly consistent with the data, the power-law index is shallower than the best fit.}
    \label{fig:mdotcompare}
\end{figure}

First, we compare the accretion rate calculated in our model for $\beta=0.3$ with the sample of \cite{hirano_one_2014} and the corresponding fit~\refeq{mdotfit}, in \reffig{mdotcompare}. Compared to the simulation results, the analytic calculation overestimates the accretion rate by a factor of a few for the least massive clouds, and underestimates the accretion rate by a similar factor for the heaviest.

\begin{figure}
    \centering
    \begin{minipage}{0.45\textwidth}
        \centering
        \includegraphics[width=0.9\textwidth]{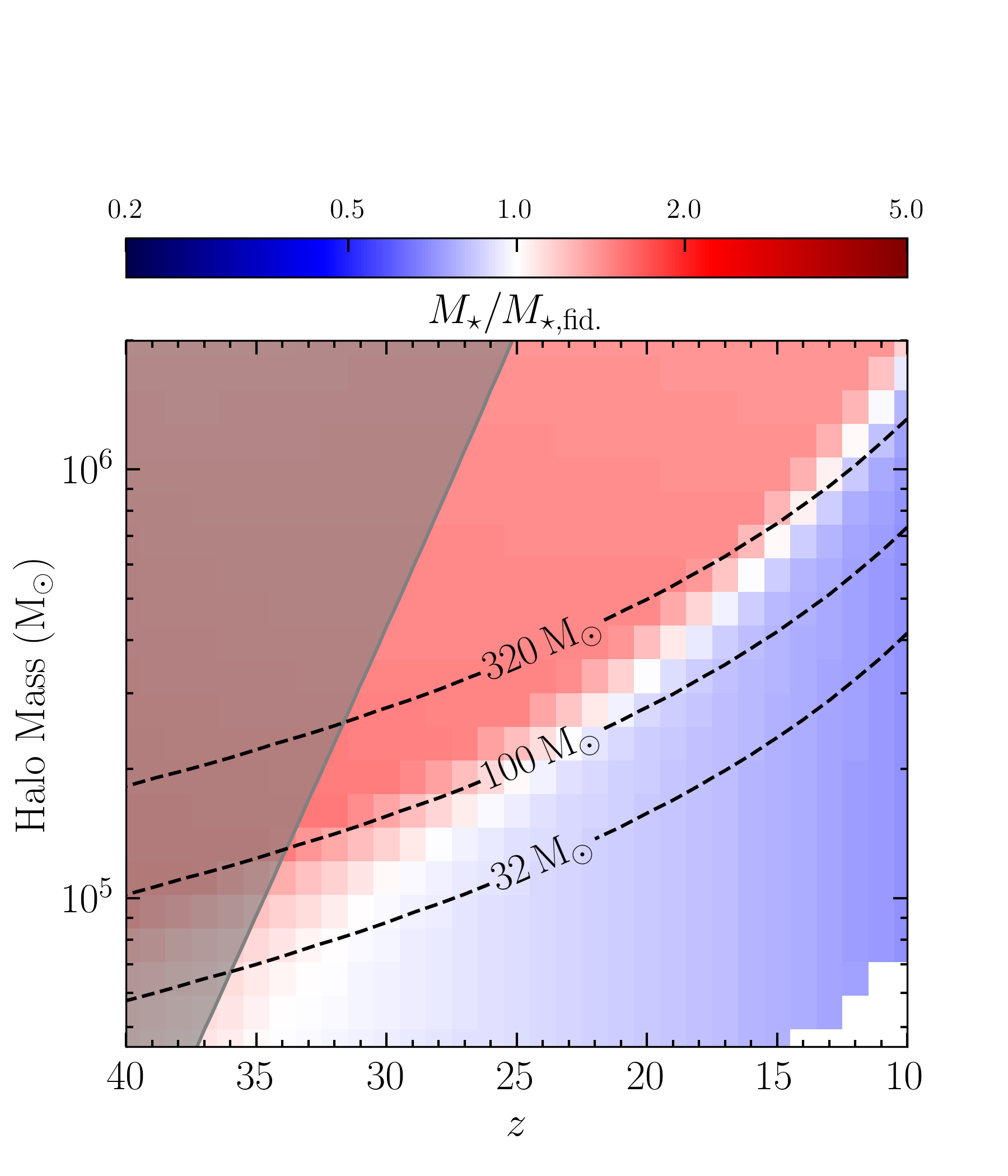}%
    \caption{The stellar mass calculated in our model using the accretion rate \refeq{mdotfit}, compared to our fiducial result (Fig.~\ref{fig:mcomp}). Also shown are contours of the fit (Eq.~\ref{eq:mstarfit}) of \cite{hirano_one_2014} to their sample. The range of stellar masses is increased by a factor of a few.}
    \label{fig:mstarmdotcompare}
    \end{minipage}\hfill
    \begin{minipage}{0.45\textwidth}
        \centering
        \includegraphics[width=0.9\textwidth]{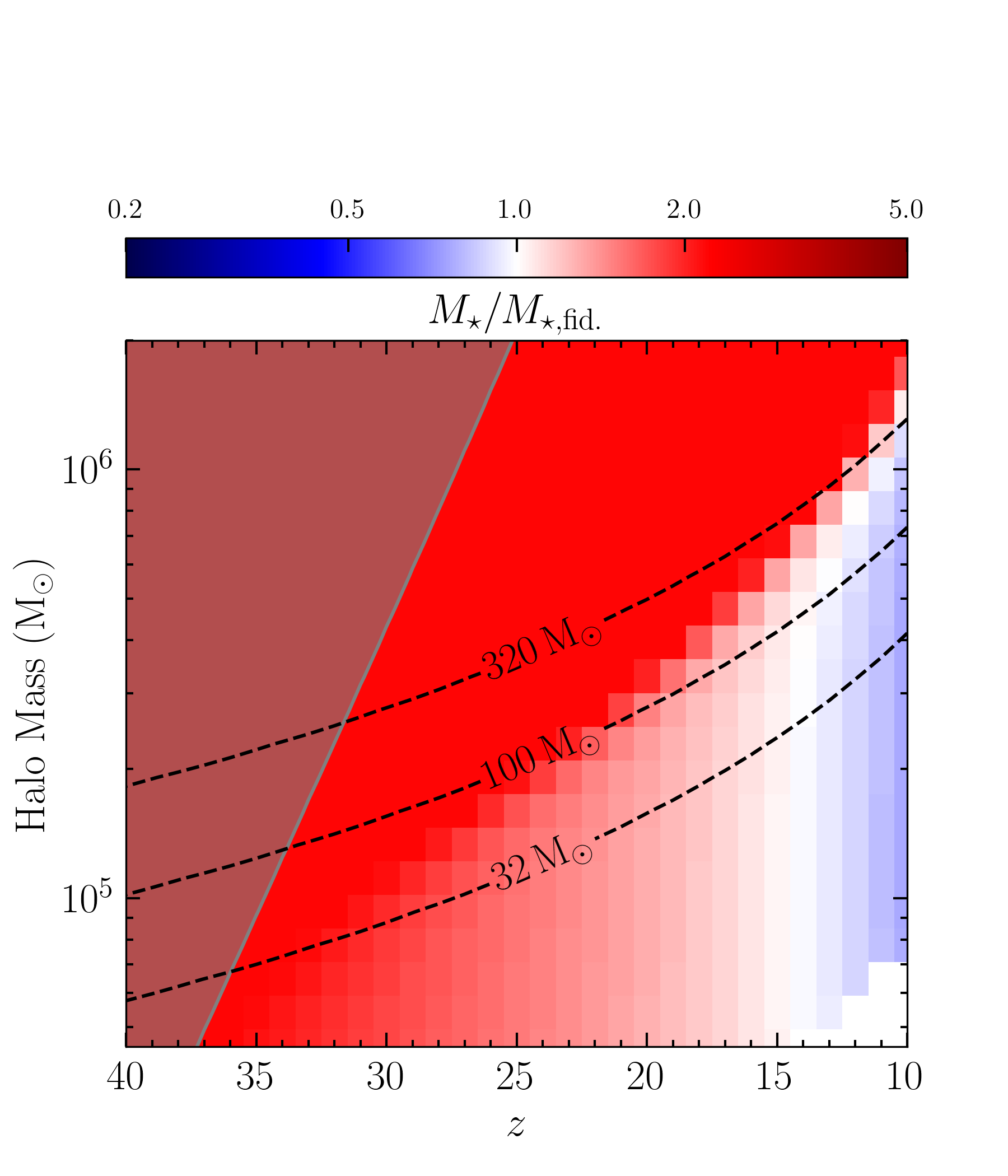}%
    \vspace{-10pt}
    \caption{The stellar mass calculated from \refeq{mfrommdot} with the accretion rate supplied by \refeq{mdotstar}. Also shown are contours of the fit (Eq.~\ref{eq:mstarfit}) of \cite{hirano_one_2014}. For most of the parameter space, the stellar mass is enhanced compared to our fiducial model. This enhancement is strongest for the most massive halos (and hence the most massive stars).}
    \label{fig:mstarfbcompare}
    \end{minipage}
\end{figure}

On the other hand, if we use our original accretion model \refeq{mdotstar} but replace our analytic model of the stellar growth and ionizing feedback with \refeq{mfrommdot}, the largest effect is a factor of a few increase in the stellar mass from the heavy, $\hto$ cooled clouds (\reffig{mstarmdotcompare}.) 

Similarly, we can use the accretion rate calculated in our model, \refeq{mdotstar}, but replace our treatment of the stellar evolution with the fit \refeq{mfrommdot}. This scenario is compared to the fiducial result in \reffig{mstarfbcompare}.

These comparisons show that both the accretion rate calculation and the treatment of the subsequent stellar evolution and feedback contribute to the discrepancies between our model and \cite{hirano_one_2014}. The largest factor arises from the stellar growth and feedback: \cite{hirano_one_2014} find that rapidly accreting stars grow to higher masses than predicted by our model. This is consistent with their finding that such stars can grow significantly even after UV breakout, likely due to self-shielding in aspherical accretion flows and early expansion of the protostar at very high accretion rates \citep[$\gtrsim 0.1\ \rm M_{\odot}/yr$,][]{hosokawa_rapidly_2012}, which are not considered in our model.

\end{document}

%% file: defs.tex
\newcommand{\refeqs}[2]{Eqs.~(\ref{eq:#1})--(\ref{eq:#2})}          
          
\renewcommand{\refeq}[1]{Eq.~(\ref{eq:#1})}          
\newcommand{\reffig}[1]{Fig.~\ref{fig:#1}}          
          
\newcommand{\reftab}[1]{Tab.~\ref{tab:#1}}
\newcommand{\refsec}[1]{Sec.~\ref{sec:#1}}

\def\apjl{Astrophys.\ J.\ Lett.}

\def\aap{Astron.\ Astrophys.}

\def\apjs{Astrophys.\ J. Supp.}

\newcommand{\be}{\begin{equation}}
\newcommand{\ee}{\end{equation}}
\newcommand{\bea}{\begin{eqnarray}}
\newcommand{\eea}{\end{eqnarray}}
 
\def\ba#1\ea{\begin{align}#1\end{align}}

\renewcommand{\O}{\mathcal{O}}

\newcommand{\T}{\mathcal{T}}
\newcommand{\Q}{\mathcal{Q}}

\renewcommand{\P}{\mathcal{P}}

\renewcommand{\v}[1]{\bm{#1}}

\newcommand{\D}{\Delta}
\newcommand{\s}{\sigma}
\renewcommand{\d}{\delta}
\renewcommand{\a}{\alpha}
\renewcommand{\o}{\omega}

\def\L{\Lambda}

\def\tt{\tilde t_{0}}

\def\Mag{\mathcal{M}}





%% file: main.bbl
\begin{thebibliography}{}
\expandafter\ifx\csname natexlab\endcsname\relax\def\natexlab#1{#1}\fi
\providecommand{\url}[1]{\href{#1}{#1}}
\providecommand{\dodoi}[1]{doi:~\href{http://doi.org/#1}{\nolinkurl{#1}}}
\providecommand{\doeprint}[1]{\href{http://ascl.net/#1}{\nolinkurl{http://ascl.net/#1}}}
\providecommand{\doarXiv}[1]{\href{https://arxiv.org/abs/#1}{\nolinkurl{https://arxiv.org/abs/#1}}}

\bibitem[{{Abel} {et~al.}(1997){Abel}, {Anninos}, {Zhang}, \&
  {Norman}}]{Abel1997}
{Abel}, T., {Anninos}, P., {Zhang}, Y., \& {Norman}, M.~L. 1997, \na, 2, 181,
  \dodoi{10.1016/S1384-1076(97)00010-9}

\bibitem[{Alvarez {et~al.}(2006)Alvarez, Bromm, \& Shapiro}]{Alvarez_2006}
Alvarez, M.~A., Bromm, V., \& Shapiro, P.~R. 2006, The Astrophysical Journal,
  639, 621, \dodoi{10.1086/499578}

\bibitem[{{Bromm}(2013)}]{Bromm:2013}
{Bromm}, V. 2013, Reports on Progress in Physics, 76, 112901,
  \dodoi{10.1088/0034-4885/76/11/112901}

\bibitem[{{Bromm} {et~al.}(2002){Bromm}, {Coppi}, \& {Larson}}]{Bromm2002}
{Bromm}, V., {Coppi}, P.~S., \& {Larson}, R.~B. 2002, \apj, 564, 23,
  \dodoi{10.1086/323947}

\bibitem[{Bromm \& Larson(2004)}]{Bromm2004}
Bromm, V., \& Larson, R.~B. 2004, Annual Review of Astronomy and Astrophysics,
  42, 79, \dodoi{10.1146/annurev.astro.42.053102.134034}

\bibitem[{{Clark} {et~al.}(2011){Clark}, {Glover}, {Smith}, {Greif}, {Klessen},
  \& {Bromm}}]{Clark2011}
{Clark}, P.~C., {Glover}, S. C.~O., {Smith}, R.~J., {et~al.} 2011, Science,
  331, 1040, \dodoi{10.1126/science.1198027}

\bibitem[{Cooke {et~al.}(2018)Cooke, Pettini, \& Steidel}]{Cooke2018}
Cooke, R.~J., Pettini, M., \& Steidel, C.~C. 2018, The Astrophysical Journal,
  855, 102, \dodoi{10.3847/1538-4357/aaab53}

\bibitem[{{Coppola} {et~al.}(2019){Coppola}, {Lique}, {Mazzia}, {Esposito}, \&
  {Kazandjian}}]{Coppola2019}
{Coppola}, C.~M., {Lique}, F., {Mazzia}, F., {Esposito}, F., \& {Kazandjian},
  M.~V. 2019, \mnras, 486, 1590, \dodoi{10.1093/mnras/stz927}

\bibitem[{{Desjacques} {et~al.}(2018){Desjacques}, {Jeong}, \&
  {Schmidt}}]{PBSreview}
{Desjacques}, V., {Jeong}, D., \& {Schmidt}, F. 2018, \physrep, 733, 1,
  \dodoi{10.1016/j.physrep.2017.12.002}

\bibitem[{{Eddington}(1926)}]{Eddington1926}
{Eddington}, A.~S. 1926, {The Internal Constitution of the Stars}

\bibitem[{Galli \& Palla(1998)}]{galli1998chemistry}
Galli, D., \& Palla, F. 1998, The Chemistry of the Early Universe.
\newblock \doarXiv{astro-ph/9803315}

\bibitem[{Galli \& Palla(2002)}]{Galli_2002}
---. 2002, Planetary and Space Science, 50, 1197–1204,
  \dodoi{10.1016/s0032-0633(02)00083-1}

\bibitem[{Gay {et~al.}(2011)Gay, Stancil, Lepp, \& Dalgarno}]{Gay_2011}
Gay, C.~D., Stancil, P.~C., Lepp, S., \& Dalgarno, A. 2011, The Astrophysical
  Journal, 737, 44, \dodoi{10.1088/0004-637X/737/1/44}

\bibitem[{Grassi {et~al.}(2014)Grassi, Bovino, Schleicher, Prieto, Seifried,
  Simoncini, \& Gianturco}]{grassi_krome_2014}
Grassi, T., Bovino, S., Schleicher, D. R.~G., {et~al.} 2014, Monthly Notices of
  the Royal Astronomical Society, 439, 2386, \dodoi{10.1093/mnras/stu114}

\bibitem[{{Greif} {et~al.}(2011){Greif}, {Springel}, {White}, {Glover},
  {Clark}, {Smith}, {Klessen}, \& {Bromm}}]{Greif2011}
{Greif}, T.~H., {Springel}, V., {White}, S. D.~M., {et~al.} 2011, \apj, 737,
  75, \dodoi{10.1088/0004-637X/737/2/75}

\bibitem[{Gurian {et~al.}(2022)Gurian, Ryan, Schon, Jeong, \&
  Shandera}]{Gurian2022}
Gurian, J., Ryan, M., Schon, S., Jeong, D., \& Shandera, S. 2022, The
  Astrophysical Journal Letters, 939, L12, \dodoi{10.3847/2041-8213/ac997c}

\bibitem[{Haemmerl{\'{e}} {et~al.}(2020)Haemmerl{\'{e}}, Mayer, Klessen,
  Hosokawa, Madau, \& Bromm}]{Haemmerl__2020}
Haemmerl{\'{e}}, L., Mayer, L., Klessen, R.~S., {et~al.} 2020, Space Science
  Reviews, 216, \dodoi{10.1007/s11214-020-00673-y}

\bibitem[{Hippert {et~al.}(2022)Hippert, Setford, Tan, Curtin, Noronha-Hostler,
  \& Yunes}]{Hippert_2022}
Hippert, M., Setford, J., Tan, H., {et~al.} 2022, Physical Review D, 106,
  \dodoi{10.1103/physrevd.106.035025}

\bibitem[{{Hirano} {et~al.}(2015){Hirano}, {Hosokawa}, {Yoshida}, {Omukai}, \&
  {Yorke}}]{Hirano2015}
{Hirano}, S., {Hosokawa}, T., {Yoshida}, N., {Omukai}, K., \& {Yorke}, H.~W.
  2015, \mnras, 448, 568, \dodoi{10.1093/mnras/stv044}

\bibitem[{Hirano {et~al.}(2014)Hirano, Hosokawa, Yoshida, Umeda, Omukai,
  Chiaki, \& Yorke}]{hirano_one_2014}
Hirano, S., Hosokawa, T., Yoshida, N., {et~al.} 2014, The Astrophysical
  Journal, 781, 60, \dodoi{10.1088/0004-637X/781/2/60}

\bibitem[{{Hirano} \& {Machida}(2022)}]{Hirano2022}
{Hirano}, S., \& {Machida}, M.~N. 2022, \apjl, 935, L16,
  \dodoi{10.3847/2041-8213/ac85e0}

\bibitem[{Hirata \& Padmanabhan(2006)}]{Hirata2006}
Hirata, C.~M., \& Padmanabhan, N. 2006, Monthly Notices of the Royal
  Astronomical Society, 372, 1175–1186,
  \dodoi{10.1111/j.1365-2966.2006.10924.x}

\bibitem[{{Hollenbach} \& {McKee}(1979)}]{Hollenbach1979}
{Hollenbach}, D., \& {McKee}, C.~F. 1979, \apjs, 41, 555,
  \dodoi{10.1086/190631}

\bibitem[{{Hosokawa} {et~al.}(2016){Hosokawa}, {Hirano}, {Kuiper}, {Yorke},
  {Omukai}, \& {Yoshida}}]{Hosokawa2016}
{Hosokawa}, T., {Hirano}, S., {Kuiper}, R., {et~al.} 2016, \apj, 824, 119,
  \dodoi{10.3847/0004-637X/824/2/119}

\bibitem[{Hosokawa {et~al.}(2012)Hosokawa, Omukai, \&
  Yorke}]{hosokawa_rapidly_2012}
Hosokawa, T., Omukai, K., \& Yorke, H.~W. 2012, The Astrophysical Journal, 756,
  93, \dodoi{10.1088/0004-637X/756/1/93}

\bibitem[{Hosokawa {et~al.}(2011)Hosokawa, Omukai, Yoshida, \&
  Yorke}]{Hosokawa2011}
Hosokawa, T., Omukai, K., Yoshida, N., \& Yorke, H.~W. 2011, Science, 334,
  1250, \dodoi{10.1126/science.1207433}

\bibitem[{{Jaura} {et~al.}(2022){Jaura}, {Glover}, {Wollenberg}, {Klessen},
  {Geen}, \& {Haemmerl{\'e}}}]{Jaura2022}
{Jaura}, O., {Glover}, S. C.~O., {Wollenberg}, K. M.~J., {et~al.} 2022, \mnras,
  512, 116, \dodoi{10.1093/mnras/stac487}

\bibitem[{{Johnson} \& {Bromm}(2006)}]{Johnson2006}
{Johnson}, J.~L., \& {Bromm}, V. 2006, \mnras, 366, 247,
  \dodoi{10.1111/j.1365-2966.2005.09846.x}

\bibitem[{Klessen \& Glover(2023)}]{klessen2023stars}
Klessen, R.~S., \& Glover, S. C.~O. 2023, The first stars: formation,
  properties, and impact.
\newblock \doarXiv{2303.12500}

\bibitem[{Latif {et~al.}(2022)Latif, Whalen, \& Khochfar}]{Latif_2022}
Latif, M.~A., Whalen, D., \& Khochfar, S. 2022, The Astrophysical Journal, 925,
  28, \dodoi{10.3847/1538-4357/ac3916}

\bibitem[{Lipovka {et~al.}(2005)Lipovka, N{\'{u} }{\~{n}}ez-L{\'{o}}pez, \&
  Avila-Reese}]{Lipovka2005}
Lipovka, A., N{\'{u} }{\~{n}}ez-L{\'{o}}pez, R., \& Avila-Reese, V. 2005,
  Monthly Notices of the Royal Astronomical Society, 361, 850,
  \dodoi{10.1111/j.1365-2966.2005.09226.x}

\bibitem[{Liu {et~al.}(2020)Liu, Meynet, \& Bromm}]{Liu2020}
Liu, B., Meynet, G., \& Bromm, V. 2020, Monthly Notices of the Royal
  Astronomical Society, 501, 643, \dodoi{10.1093/mnras/staa3671}

\bibitem[{Magnus {et~al.}(2023)Magnus, Smith, Khochfar, O'Shea, Wise, Norman,
  \& Turk}]{magnus2023formation}
Magnus, L.~C., Smith, B.~D., Khochfar, S., {et~al.} 2023, Formation Channels
  for Population III Stars at Cosmic Dawn.
\newblock \doarXiv{2307.03521}

\bibitem[{McKee \& Tan(2008)}]{mckee_formation_2008}
McKee, C.~F., \& Tan, J.~C. 2008, The Astrophysical Journal, 681, 771,
  \dodoi{10.1086/587434}

\bibitem[{Nakazato {et~al.}(2022)Nakazato, Chiaki, Yoshida, Naoz, Lake, \&
  Chiou}]{Nakazato_2022}
Nakazato, Y., Chiaki, G., Yoshida, N., {et~al.} 2022, The Astrophysical Journal
  Letters, 927, L12, \dodoi{10.3847/2041-8213/ac573e}

\bibitem[{Nebrin {et~al.}(2023)Nebrin, Giri, \& Mellema}]{Nebrin_2023}
Nebrin, O., Giri, S.~K., \& Mellema, G. 2023, Monthly Notices of the Royal
  Astronomical Society, 524, 2290, \dodoi{10.1093/mnras/stad1852}

\bibitem[{Omukai \& Nishi(1998)}]{Omukai_1998}
Omukai, K., \& Nishi, R. 1998, The Astrophysical Journal, 508, 141,
  \dodoi{10.1086/306395}

\bibitem[{{Park} {et~al.}(2023{\natexlab{a}}){Park}, {Ricotti}, \&
  {Sugimura}}]{Park2023}
{Park}, J., {Ricotti}, M., \& {Sugimura}, K. 2023{\natexlab{a}}, arXiv
  e-prints, arXiv:2307.14562, \dodoi{10.48550/arXiv.2307.14562}

\bibitem[{{Park} {et~al.}(2023{\natexlab{b}}){Park}, {Ricotti}, \&
  {Sugimura}}]{Park2023a}
---. 2023{\natexlab{b}}, \mnras, 521, 5334, \dodoi{10.1093/mnras/stad895}

\bibitem[{{Press} \& {Schechter}(1974)}]{Press1974}
{Press}, W.~H., \& {Schechter}, P. 1974, \apj, 187, 425, \dodoi{10.1086/152650}

\bibitem[{{Rees}(1976)}]{Rees1976}
{Rees}, M.~J. 1976, \mnras, 176, 483, \dodoi{10.1093/mnras/176.3.483}

\bibitem[{Ripamonti(2007)}]{Ripamonti_2007}
Ripamonti, E. 2007, Monthly Notices of the Royal Astronomical Society, 376,
  709, \dodoi{10.1111/j.1365-2966.2007.11460.x}

\bibitem[{Ryan \& Radice(2022)}]{Ryan2022}
Ryan, M., \& Radice, D. 2022, Phys. Rev. D, 105, 115034,
  \dodoi{10.1103/PhysRevD.105.115034}

\bibitem[{Seager {et~al.}(1999)Seager, Sasselov, \& Scott}]{Seager_1999}
Seager, S., Sasselov, D.~D., \& Scott, D. 1999, The Astrophysical Journal, 523,
  L1–L5, \dodoi{10.1086/312250}

\bibitem[{{Shakura} \& {Sunyaev}(1973)}]{Shakura1973}
{Shakura}, N.~I., \& {Sunyaev}, R.~A. 1973, \aap, 24, 337

\bibitem[{Shandera {et~al.}(2018)Shandera, Jeong, \& Gebhardt}]{Shandera_2018}
Shandera, S., Jeong, D., \& Gebhardt, H. S.~G. 2018, Physical Review Letters,
  120, \dodoi{10.1103/physrevlett.120.241102}

\bibitem[{Sharda {et~al.}(2020)Sharda, Federrath, \& Krumholz}]{Sharda_2020}
Sharda, P., Federrath, C., \& Krumholz, M.~R. 2020, Monthly Notices of the
  Royal Astronomical Society, 497, 336, \dodoi{10.1093/mnras/staa1926}

\bibitem[{Sharda {et~al.}(2021)Sharda, Federrath, Krumholz, \&
  Schleicher}]{Sharda_2021}
Sharda, P., Federrath, C., Krumholz, M.~R., \& Schleicher, D. R.~G. 2021,
  Monthly Notices of the Royal Astronomical Society, 503, 2014,
  \dodoi{10.1093/mnras/stab531}

\bibitem[{Sheth {et~al.}(2001)Sheth, Mo, \& Tormen}]{Sheth2001}
Sheth, R.~K., Mo, H.~J., \& Tormen, G. 2001, Monthly Notices of the Royal
  Astronomical Society, 323, 1, \dodoi{10.1046/j.1365-8711.2001.04006.x}

\bibitem[{Shu {et~al.}(2002)Shu, Lizano, Galli, Canto, \& Laughlin}]{Shu_2002}
Shu, F.~H., Lizano, S., Galli, D., Canto, J., \& Laughlin, G. 2002, The
  Astrophysical Journal, 580, 969, \dodoi{10.1086/343859}

\bibitem[{Singh {et~al.}(2021)Singh, Ryan, Magee, Akhter, Shandera, Jeong, \&
  Hanna}]{Singh_2021}
Singh, D., Ryan, M., Magee, R., {et~al.} 2021, Physical Review D, 104,
  \dodoi{10.1103/physrevd.104.044015}

\bibitem[{Stacy \& Bromm(2013)}]{stacy_constraining_2013}
Stacy, A., \& Bromm, V. 2013, Monthly Notices of the Royal Astronomical
  Society, 433, 1094, \dodoi{10.1093/mnras/stt789}

\bibitem[{Stacy {et~al.}(2016)Stacy, Bromm, \& Lee}]{Stacy2016}
Stacy, A., Bromm, V., \& Lee, A.~T. 2016, Monthly Notices of the Royal
  Astronomical Society, 462, 1307, \dodoi{10.1093/mnras/stw1728}

\bibitem[{{Stacy} {et~al.}(2010){Stacy}, {Greif}, \& {Bromm}}]{Stacy2010}
{Stacy}, A., {Greif}, T.~H., \& {Bromm}, V. 2010, \mnras, 403, 45,
  \dodoi{10.1111/j.1365-2966.2009.16113.x}

\bibitem[{{Stahler} {et~al.}(1986){Stahler}, {Palla}, \&
  {Salpeter}}]{Stahler1986}
{Stahler}, S.~W., {Palla}, F., \& {Salpeter}, E.~E. 1986, \apj, 308, 697,
  \dodoi{10.1086/164542}

\bibitem[{{Steigman}(2007)}]{BBNreview:2007}
{Steigman}, G. 2007, Annual Review of Nuclear and Particle Science, 57, 463,
  \dodoi{10.1146/annurev.nucl.56.080805.140437}

\bibitem[{{Sugimura} {et~al.}(2020){Sugimura}, {Matsumoto}, {Hosokawa},
  {Hirano}, \& {Omukai}}]{Sugimura2020}
{Sugimura}, K., {Matsumoto}, T., {Hosokawa}, T., {Hirano}, S., \& {Omukai}, K.
  2020, \apjl, 892, L14, \dodoi{10.3847/2041-8213/ab7d37}

\bibitem[{{Sugimura} {et~al.}(2023){Sugimura}, {Matsumoto}, {Hosokawa},
  {Hirano}, \& {Omukai}}]{Sugimura2023}
---. 2023, arXiv e-prints, arXiv:2307.15108, \dodoi{10.48550/arXiv.2307.15108}

\bibitem[{Susa(2019)}]{susa_merge_2019}
Susa, H. 2019, The Astrophysical Journal, 877, 99,
  \dodoi{10.3847/1538-4357/ab1b6f}

\bibitem[{{Susa} {et~al.}(2014){Susa}, {Hasegawa}, \& {Tominaga}}]{Susa2014}
{Susa}, H., {Hasegawa}, K., \& {Tominaga}, N. 2014, \apj, 792, 32,
  \dodoi{10.1088/0004-637X/792/1/32}

\bibitem[{Tegmark {et~al.}(1997)Tegmark, Silk, Rees, Blanchard, Abel, \&
  Palla}]{Tegmark1997}
Tegmark, M., Silk, J., Rees, M.~J., {et~al.} 1997, The Astrophysical Journal,
  474, 1, \dodoi{10.1086/303434}

\bibitem[{{Turk} {et~al.}(2009){Turk}, {Abel}, \& {O'Shea}}]{Turk2009}
{Turk}, M.~J., {Abel}, T., \& {O'Shea}, B. 2009, Science, 325, 601,
  \dodoi{10.1126/science.1173540}

\bibitem[{{Yoshida} {et~al.}(2006){Yoshida}, {Omukai}, {Hernquist}, \&
  {Abel}}]{Yoshida/etal:2006}
{Yoshida}, N., {Omukai}, K., {Hernquist}, L., \& {Abel}, T. 2006, \apj, 652, 6,
  \dodoi{10.1086/507978}

\end{thebibliography}
